\documentclass[11pt]{article}
\RequirePackage{algorithm,algorithmic,graphicx,amssymb,amsmath,color,pifont,enumerate}
\RequirePackage{fullpage}

\renewcommand{\deg}{\mathsf{deg}}
\newcommand{\lowdeg}{\mathsf{d_{low}}}

\newtheorem{invariant}{Invariant}

\newcommand{\evalf}{\texttt{handle-free}}

\newcommand{\ins}{\texttt{handle-insertion}}
\newcommand{\del}{\texttt{handle-deletion}}
\newcommand{\detr}{\texttt{deterministic-settle}}
\newcommand{\rand}{\texttt{random-settle}}

\newcommand{\mate}{\texttt{mate}}

\newcommand{\dout}{d_{\texttt{out}}}

\newcommand{\setlvl}{\texttt{set-level}}

\newcommand {\ignore} [1] {}

\def\denseformat{
\setlength{\textheight}{9in}
\setlength{\textwidth}{6.9in}
\setlength{\evensidemargin}{-0.2in}
\setlength{\oddsidemargin}{-0.2in}
\setlength{\headsep}{10pt}
\setlength{\topmargin}{-0.3in}
\setlength{\columnsep}{0.375in}
\setlength{\itemsep}{0pt}
}

\newtheorem{theorem}{Theorem}[section]

\newtheorem{claim}[theorem]{Claim}
\newtheorem{lemma}[theorem]{Lemma}

\newtheorem{corollary}[theorem]{Corollary}

\newtheorem{comment}[theorem]{Comment}

\newtheorem{observation}[theorem]{Observation}

\def\boldhead#1:{\par\vskip 7pt\noindent{\bf #1:}\hskip 10pt}
\def\ithead#1:{\par\vskip 7pt\noindent{\it #1:}\hskip 10pt}

\def\inline#1:{\par\vskip 7pt\noindent{\bf #1:}\hskip 10pt}
\def\midinline#1:{\par\noindent{\bf #1:}\hskip 10pt}
\def\dnsinline#1:{\par\vskip -7pt\noindent{\bf #1:}\hskip 10pt}
\def\ddnsinline#1:{\newline{\bf #1:}\hskip 10pt}
\def\largeinline#1:{\par\vskip 7pt\noindent{\large\bf #1:}\hskip 10pt}
\long\def\comment #1\commentend{}
\long\def\commhide #1\commhideend{}
\long\def\commfull #1\commend{#1}
\long\def\commabs #1\commenda{}
\long\def\commtim #1\commendt{#1}
\long\def\commb #1\commbend{}
\long\def\commedit #1\commeditend{} 

\long\def\commB #1\commBend{}       

\long\def\commex #1\commexend{}     

\long\def\commsiena #1\commsienaend{}  
                                         
\long\def\commBI #1\commBIend{}  
                                         
\long\def\CProof #1\CQED{}

\def\blackslug{\hbox{\hskip 1pt \vrule width 4pt height 8pt
    depth 1.5pt \hskip 1pt}}
\def\QED{\quad\blackslug\lower 8.5pt\null\par}
\def\inQED{\quad\quad\blackslug}

\long\def\PPP#1{\noindent{\bf Proof:}{ #1}{\quad\blackslug\lower 8.5pt\null}}

\long\def\denspar #1\densend
{#1}

\newif\ifnotesw\noteswtrue
   {\ifnotesw\marginpar[\hfill\(\top\)]{\(\top\)}\fi}%
   {\ifnotesw\marginpar[\hfill\(\bot\)]{\(\bot\)}\fi}

\newcommand{\mnote}[1]%
    {\ifnotesw\marginpar%
        [{\scriptsize\it\begin{minipage}[t]{\marginparwidth}
        \raggedleft#1%
                        \end{minipage}}]%
        {\scriptsize\it\begin{minipage}[t]{\marginparwidth}
        \raggedright#1%
                        \end{minipage}}%
    \fi}

\def\cE{{\cal E}}

\def\cI{{\cal I}}

\def\cM{{\cal M}}
\def\cN{{\cal N}}
\def\cO{{\cal O}}

\def\MathF{\hbox{\rm I\kern-2pt F}}
\def\MathP{\hbox{\rm I\kern-2pt P}}
\def\MathR{\hbox{\rm I\kern-2pt R}}
\def\MathZ{\hbox{\sf Z\kern-4pt Z}}
\def\MathN{\hbox{\rm I\kern-2pt I\kern-3.1pt N}}
\def\MathC{\hbox{\rm \kern0.7pt\raise0.8pt\hbox{\footnotesize I}
\kern-4.2pt C}}
\def\MathQ{\hbox{\rm I\kern-6pt Q}}

\def\MathE{\hbox{{\rm I}\hskip -2pt {\rm E}}} 

\newsavebox{\ttop}\newsavebox{\bbot}

\def\eps{\epsilon}

\newcommand{\Prob}{\MathP}
\newcommand{\Expect}{\MathE}

\denseformat

\long\def\commabs #1\commabsend{}
\newcommand{\orient}[2]{#1 \rightarrow #2}
\begin{document}

\title{Fully Dynamic Maximal Matching in Constant Update Time}
\author{
Shay Solomon \thanks{School of Computer Science, Tel Aviv University, Tel Aviv 69978, Israel.
E-mail: {\tt solo.shay@gmail.com}.}}

\date{\empty}

\begin{titlepage}
\def\thepage{}
\maketitle

\begin{abstract}
Baswana, Gupta and Sen [FOCS'11] showed that fully dynamic maximal matching can be maintained in general graphs with logarithmic amortized update time.
More specifically, starting from an empty graph on $n$ fixed vertices, they devised a randomized algorithm for maintaining maximal matching
over any sequence of $t$ edge insertions and deletions with a total runtime of $O(t \log n)$ in expectation and $O(t \log n + n \log^2 n)$ with high probability.
Whether or not this runtime bound can be improved towards $O(t)$ has remained an important open problem.
Despite significant research efforts, this question has resisted numerous attempts at resolution even for basic graph families such as forests.

In this paper, we resolve the question in the affirmative, by presenting a randomized algorithm for maintaining maximal matching in general graphs with \emph{constant} amortized update time. The optimal   runtime bound $O(t)$ of our algorithm holds both in expectation and with high probability.

As an immediate corollary, we can maintain 2-approximate vertex cover with constant amortized update time.
This result is essentially the best one can hope for (under the unique games conjecture) in the context of dynamic approximate vertex cover, culminating a long line of research.

Our algorithm builds on Baswana et al.'s algorithm, but is inherently different and arguably simpler.
As an implication of our simplified approach, the space usage of our algorithm is linear in the (dynamic) graph size,
while the space usage of Baswana et al.'s algorithm is always at least $\Omega(n \log n)$.

Finally, we present applications to approximate weighted matchings and to distributed networks.  
\end{abstract}

\end{titlepage}

\pagenumbering {arabic}

\section{Introduction} \label{sec:intro}
{\bf 1.1~ Dynamic maximal matching.~}
For any graph $G = (V,E)$, with $n = |V|, m = |E|$, an (inclusion-wise) \emph{maximal matching} $\cM = \cM(G)$ can be computed in time $O(n+m)$ via a na\"{\i}ve greedy algorithm.
A fundamental challenge is to efficiently maintain a maximal matching in a \emph{fully dynamic setting}.

Starting from an empty graph $G_0$ on $n$ fixed vertices, at each time step $i$ a single edge $(u,v)$ is either inserted to the graph $G_{i-1}$ or deleted from it, resulting in graph $G_i$,
and the dynamic algorithm should update the maintained matching $\cM = \cM(G_{i-1})$   to preserve maximality w.r.t.\ the new graph $G_i$.
The holy grail is that the total runtime of the algorithm would be linear in the total number $t$ of update steps.
Put in other words, if the total runtime is $T$,   the \emph{amortized} update time $T/t$ should be constant.

There is a na\"{\i}ve deterministic algorithm for maintaining a maximal matching with an update time of $O(n)$.\footnote{In fact, the na\"{\i}ve update time bound holds in the \emph{worst case}.
In this paper we focus on amortized time bounds, and so in some of the literature survey that follows we do not stress the distinction between amortized versus worst-case bounds.}
In 1993, Ivkovi\'{c} and Lloyd \cite{IL93} devised a deterministic algorithm with update time $O((n+m)^{\frac{\sqrt{2}}{2}})$,
which, despite being quite involved, is inferior to the na\"{\i}ve algorithm in the regime $m = \omega(n^{\sqrt{2}})$.
The state-of-the-art deterministic update time, by Neiman and this author \cite{NS13},
is $O(\sqrt{m})$.

Baswana, Gupta and Sen \cite{BGS11} used randomization to obtain an exponential improvement in the update time, under the oblivious adversarial model.\footnote{The \emph{oblivious adversarial model} is a standard model, which has been   used for analyzing   randomized data-structures such as universal hashing \cite{CW77b}
and dynamic connectivity \cite{KKM13}.
The model allows the adversary to know all the edges in the graph and their arrival order, as well as
the algorithm to be used. However, the adversary is not aware of the random bits used by the algorithm,
and so cannot choose updates adaptively in response to the randomly guided choices of the algorithm.}
Specifically,  they devised a randomized algorithm for maintaining a maximal matching
over any sequence of $t$ edge insertions and deletions with a total runtime of $O(t \log n)$ in expectation and $O(t \log n + n \log^2 n)$ with high probability (w.h.p.).
In other words, the (amortized) update time of their algorithm is $O(\log n)$ in expectation and $O(\log n + \frac{n \log^2 n}{t})$ w.h.p.
(For $t = o(n \log n)$, the high probability bound becomes super-logarithmic; e.g., for $t = \Theta(n)$, it is $O(\log^2 n)$.)

Whether or not the update time of \cite{BGS11} can be improved towards constant has remained an important open problem.
Some progress towards its resolution was made for uniformly sparse graphs, such as forests, planar graphs and graphs excluding fixed minors \cite{NS13,HTZ14,KKPS14}.
However, the state-of-the-art update time in such graphs is $O(\sqrt{\log n})$ \cite{HTZ14}, even in forests, which is still far from constant.
\ignore{
{The \emph{arboricity} $\alpha(G)$ of a graph $G$ is the minimum number of edge-disjoint forests into which it can be partitioned,
and it is close to the density of its densest subgraph. (For any $m$-edge graph, its arboricity ranges between 1 and $\sqrt{m}$.)}
For constant arboricity graphs (e.g., planar graphs and graphs excluding fixed minors), Neiman and this author \cite{NS13} devised a deterministic algorithm with update time  $O(\log n/\log \log n)$. This update time was improved to $O(\sqrt{\log n})$ by He, Tang and Zeh \cite{HTZ14}.
The general update time of \cite{HTZ14} is $O(\alpha + \sqrt{\alpha \log n})$,
provided that the dynamic graph has arboricity at most $\alpha$ at all times. The update time bounds  of \cite{NS13,HTZ14} are amortized,
but similar (though inferior) worst-case bounds were given in \cite{KKPS14}.
}
\vspace{4pt}
\\
\noindent
{\bf Our contribution.~}
We resolve this basic question in the affirmative, by presenting a randomized algorithm for maintaining maximal matching in \emph{general graphs} with \emph{constant}   update time. The optimal runtime bound $O(t)$ of our algorithm holds both in expectation and w.h.p. (See Table \ref{tab1} in App.\ \ref{app:tab}.)


Our algorithm builds on Baswana et al.'s algorithm \cite{BGS11}, but is \emph{inherently different}.
Also, it is arguably much simpler,  both conceptually and technically.
This simplification is, in our opinion, an important contribution by itself. As a practical implication of our simplified approach,
we   implement the algorithm using \emph{optimal} space $O(n+m)$, where $m$ stands for the dynamic number of edges in the graph.
This should be contrasted to the space usage $O(n\log n+m)$ of \cite{BGS11},
which is suboptimal when $m = o(n \log n)$. In particular, whenever $m = O(n)$ (which is always the case in uniformly sparse graphs, e.g., planar graphs), this leads to a logarithmic space improvement.
Although   simpler than \cite{BGS11}, our algorithm is far from being simple -- indeed, it is   tricky and sophisticated.
(See Sections 1.4 and 2 for details.)
\vspace{9pt}
\\
\noindent
{\bf 1.2~ Dynamic approximate matching and vertex cover.~}
On static graphs, the classic maximum cardinality matching (MCM) algorithms \cite{HK73,MV80,Vaz12} run in $O(m \sqrt{n})$ time.
For dynamic MCMs, Sankowski \cite{Sank07} devised a randomized algorithm with update time $O(n^{1.495})$.\footnote{Abboud and Vassilevska Williams \cite{AW14} proved conditional lower bounds for dynamic MCMs, showing that the update time must be
$\Omega(m^{\eps})$ under
conjectures concerning the complexity of triangle detection, combinatorial boolean matrix multiplication and 3SUM, where $\eps$ is a constant
depending on the specific conjecture.
For example, under the 3SUM conjecture, it was shown in \cite{AW14} that $\eps \ge 1/3$; moreover, under this conjecture, a stronger bound of $\eps \ge 1/2 -o(1)$ was given in \cite{KPP16}.}
On the other hand, finding a minimum cardinality  vertex cover (MCVC) is NP-hard.
Moreover, under the unique games conjecture (UGC), the MCVC cannot be efficiently approximated within any factor better than 2 \cite{KR08}.
Thus, while the ultimate goal in the context of dynamic MCMs is to efficiently maintain an exact MCM (or maybe a $(1+\eps)$-MCM), the analogous goal for dynamic MCVCs is to maintain a 2-MCVC, where $t$-MCM (resp., $t$-MCVC) is a shortcut for $t$-approximate MCM (resp., $t$-approximate MCVC), for  any $t \ge 1$.
Despite the inherent difference between these   problems, they remain closely related as their LP-relaxations are duals of each other.
Note also that (1) the MCM and MCVC are of the same size up to a factor of 2;
(2) the set of matched vertices in a maximal matching is a 2-MCVC; (3) a maximal matching is a 2-MCM.




In view of the difficulty of dynamically maintaining \emph{exact} MCMs and MCVCs,
there has been a growing interest in dynamic approximate matching and vertex cover since
the pioneering work of Onak and Rubinfeld \cite{OR10}, who presented a randomized algorithm
for maintaining $c$-MCM and $c$-MCVC, where $c$ is a large unspecified constant.
The update time of the algorithm of \cite{OR10} is  w.h.p.\ $O(\log^2 n)$.

The dynamic maximal matching algorithms discussed in Section 1.1 provide 2-MCMs. Moreover, they imply dynamic algorithms for maintaining
 2-MCVCs with the same (up to constants) update time.
In particular, Baswana et al.'s algorithm \cite{BGS11} can be used to maintain 2-MCM and 2-MCVC with update time
$O(\log n)$ in expectation and $O(\log n + \frac{n \log^2 n}{t})$ w.h.p., significantly improving the result of \cite{OR10}.

There are many other works on dynamic approximate MCMs and MCVCs; we briefly mention the state-of-the-art results:
\cite{BHI15} presented a dynamic algorithm for $(2+\eps)$-MCVCs with  update time $O(\log n \cdot \eps^{-2})$;
\cite{GP13} gave an algorithm for $(1+\eps)$-MCMs with update time $O(\sqrt{m} \cdot \eps^{-2})$;
\cite{BS15,BS16} devised dynamic algorithms for $(3/2+\eps)$-MCMs in general graphs with update time $O({m}^{1/4} \cdot \eps^{-2.5})$.
For uniformly sparse graphs,
\cite{PS16} presented algorithms for maintaining $(1+\eps)$-MCMs and $(2+\eps)$-MCVCs with update times
that depend on the density (or \emph{arboricity}) of the graph and $\eps$.
(See Tables \ref{tab2} and \ref{tab3} in App.\ \ref{app:tab}.)
\vspace{4pt}
\\
\noindent
{\bf Our contribution.~}
Our dynamic maximal matching algorithm can be used to maintain 2-MCM  and 2-MCVC  in general graphs with update time that is bounded both in expectation and w.h.p.\ by constant.
Under the UGC, a $(2-\eps)$-MCVC cannot be efficiently maintained, for any $\eps > 0$.
Hence, our dynamic 2-MCVC  algorithm is essentially the best one can hope for,
culminating a long line of research.
\vspace{7pt}
\\
\noindent
{\bf 1.3~ Additional applications.~} Static and dynamic algorithms for exact and approximate maximum weighted matchings have been extensively studied, both in centralized and distributed networks
(see, e.g., \cite{LPP08,LPR09,Pettie12,ABGS12,GP13,DP14,Gupta14}).
In particular, Anand et al.\ \cite{ABGS12} gave a randomized algorithm for maintaining an 8-approximate maximum weight matching (8-MWM)
in general $n$-vertex weighted graphs with expected update time $O(\log n \log \Delta)$, where $\Delta$ is the ratio between
the maximum and minimum edge weights in the graph. The algorithm of \cite{ABGS12} employs the dynamic maximal matching algorithm of \cite{BGS11} as a black-box.
By plugging our improved algorithm, we shave a factor of $\log n$ from the update time, obtaining an algorithm for maintaining an 8-MWM
in general weighted graphs with expected update time $O(\log \Delta)$.

Another application of our dynamic maximal matching algorithm is to \emph{distributed networks}.
In a static distributed setting all processors wake up simultaneously,
and computation proceeds in fault-free synchronous rounds during which every processor exchanges messages of size $O(\log n)$.\footnote{We consider the standard $\mathcal{CONGEST}$ model (cf.~\cite{PelB00}), which captures the essence of spatial locality and congestion.}
In a dynamic distributed setting, however, upon the insertion or deletion of an edge $e=(u,v)$, only the affected vertices ($u$ and $v$) are woken up.
In this setting, the goal is to optimize (1) the number of communication rounds, and (2) the number of messages,
needed for repairing the solution per update operation,  over a worst-case sequence of update operations.
Following an edge update, $O(1)$ communication rounds trivially suffice for updating a maximal matching (and thus 2-MCM and 2-MCVC).
However, the number of messages sent per update may be as high as $O(n)$.
Our dynamic maximal matching algorithm can be easily distributed, so that the average number of messages sent per update is a small constant.

A more detailed discussion on the applications of our main result is deferred to App.\ \ref{app:app}.
\vspace{7pt}
\\
\noindent
{\bf 1.4~ Our and Previous Techniques.}
At the core of Baswana et al.\ (hereafter, \emph{BGS}) algorithm \cite{BGS11} is a complex bucketing scheme,
where each vertex is assigned   a unique \emph{level} among $\{-1,0,\ldots, \log n\}$ according to some stringent criteria.
(See Sect.\ \ref{sec:tech} for more details.)
The BGS algorithm constantly and persistently changes the level of vertices.
Changing the level of a single vertex may trigger a sequence of changes to  multiple lower level vertices, referred to in \cite{BGS11} as a \emph{fading wave}.
The cost of   a fading wave may be linear in the level of the vertex initiated it, thus the update time is $O(\log n)$. 
This is not the only logarithmic-time bottleneck incurred by the fading wave mechanism.
Indeed, a central ingredient of this mechanism  determines the ``right'' level to which a vertex should move.
To this end, an explicit counter $\phi_v[j]$ in maintained for each vertex $v$ and any level $j \in \{-1,0,\ldots, \log n\}$.
Following a single update  on vertices $u$ and $v$, the BGS algorithm has to update a logarithmic number of counters for $u$ and $v$.
Moreover, finding the right level to which a vertex should move using these counters
cannot be done in sublogarithmic time (via a binary search on any other search method) due to the special nature of these counters.
On top of these logarithmic-time bottlenecks, the fading wave mechanism of BGS is highly intricate, in terms of both the implementation and the analysis.

To break the logarithmic-time barrier, one has to take a significantly different approach, which circumvents the fading wave mechanism,
and more importantly, the usage of the counters $\phi_v[j]$.
Although our approach builds on the BGS scheme, it borrows mainly from its simpler ideas, successfully avoiding the use of the more complicated ones.
Thus, despite the resemblance between the two approaches, our algorithm deviates significantly from the BGS scheme.
We too use a bucketing scheme with $O(\log n)$ levels.
However, to remove the dependency on the number of levels from the update time,
our level-maintenance scheme cannot follow stringent criteria as in BGS.
In particular, we use a ``lazy'' approach, which changes the level of a vertex only when necessary.
When this happens, it essentially means that the vertex will ``rise'' to a higher level.
As opposed to the BGS algorithm that uses counters to determine the ``right'' level, our algorithm figures out the ``right'' level ``on the fly''
via a certain ``level-rising mechanism''. (The ``right'' level computed by our algorithm is different than that computed by the BGS algorithm.)

The ``level-rising mechanism'' is a central ingredient in our algorithm, and is obtained by combining ideas from BGS with numerous novel fundamental ideas. (See Sections 2 and 3 for details.)
We anticipate that our level-rising mechanism will be applicable to additional dynamic graph problems under the oblivious adversarial model,
where some subgraph structure has to be maintained.
(At the very least, it should give rise to improved algorithms for maintaining dynamic $(2-\eps)$-MCMs.
In fact, we have recently achieved some new results in this context, but they lie outside the scope of the current paper.)


\vspace{-4pt}
\section{A High-Level Technical Overview} \label{sec:tech}
\vspace{-4pt}
{\bf The Basics.}
Let $\cM$ be a maximal matching for the graph $G = (V,E)$ at a certain update step.
Any edge of $\cM$ (resp., $E \setminus \cM$) is called \emph{matched} (resp., \emph{unmatched}).
A vertex incident on a matched edge is called \emph{matched}, and the other endpoint is its \emph{mate}; otherwise it is   \emph{free}.
Consider the following na\"{\i}ve algorithm for maintaining a maximal matching.
Following an edge update $e=(u,v)$, if $e$ is inserted to the graph and its two endpoints are free,   the algorithm adds $e$ to $\cM$.
If   $e$ is deleted from the graph and $e \in \cM$,
the algorithm first removes $e$ from $\cM$.
Next, let $z \in \{u,v\}$;
if $z$ is free, the algorithm scans \emph{all} neighbors of $z$ looking for a free vertex.
If a free neighbor $w$ of $z$ is found,   the edge $(z,w)$ is added to $\cM$.
Clearly, the update time is constant except for the case that a matched edge gets deleted from the graph, and then
the update time is $O(\deg(u)+\deg(v))$, which may be as high as $O(n)$, even in forests.

Following an edge deletion $(u,v)$, the na\"{\i}ve algorithm matches $z \in \{u,v\}$ with an \emph{arbitrary} free neighbor $w$.
What if we match $z$ with a \emph{random} neighbor $w$?
Under the oblivious adversarial model, the expected number of edges incident on $z$ that are deleted from the graph before deleting edge $(z,w)$
is $\deg(z)/2$.
Hence, the expected \emph{amortized} cost of the deletion of edge $(z,w)$ should be $\frac{O(\deg(z) + \deg(w))}{\deg(z)/2}$.
If $\deg(w) = O(\deg(z))$, this cost is constant; however, in the general case it may be as high as  $O(n)$.

Instead of choosing $w$ randomly among all neighbors of $z$, one may restrict the attention to the low-degree neighbors of $z$, i.e., with degree $O(\deg(z))$.
Denoting by $\lowdeg(z)$ the number of such neighbors,   the amortized cost becomes $\frac{O(\deg(z))}{\lowdeg(z)/2}$.
However, if $\lowdeg(z) \ll \deg(z)$, this cost may again be too high.

Moreover, even under the optimistic assumption that $\deg(z) = O(\lowdeg(z))$, there is still a fundamental problem:
A random neighbor of $z$ (of low degree or not) may be matched. To match $z$ with a neighbor $w$ that is matched to say $w'$, we must first delete edge $(w,w')$
from $\cM$. However, the cost of such \emph{induced deletions} cannot be bounded in the amortized sense, as the adversarial argument does not hold.

\vspace{4pt}
\noindent
{\bf Baswana et al.'s scheme.~}
The BGS scheme \cite{BGS11} can be described as follows. Each edge is assigned an \emph{orientation}, yielding a directed graph,
and the vertices are given \emph{preferences} according to their out-degrees.  
The edge's orientation also induces \emph{responsibility} for one of its endpoints:
If edge $(u,v)$ is oriented as ${\orient u v}$, then it is within $u$'s responsibility to notify $v$ about any change concerning $u$.
Therefore, at any point in time, each vertex knows the updated information of its incoming neighbors.
In order to obtain a complete information of all its neighbors,
the vertex just needs to scan its outgoing neighbors.
Thus, the cost of finding a free neighbor for vertex $z$ reduces from $O(\deg(z))$ to $O(\dout(z))$, where $\dout(z)$ is $z$'s out-degree.
Assuming we choose the mate of a vertex uniformly at random among all its outgoing neighbors
(and disregarding the issue of induced deletions),
the amortized cost should reduce from $\frac{O(\deg(z) + \deg(w))}{\deg(z)/2}$ to $\frac{O(\dout(z) + \dout(w))}{\dout(z)/2}$.
Similarly to before, if we could guarantee that $\dout(w) = O(\dout(z))$, the amortized cost would be further reduced to constant.
To this end, ideally, we should have each edge oriented towards the lower out-degree endpoint.
Alas, a single edge update may trigger a long cascade of edge flips.


The strategy of \cite{BGS11} is to maintain for each vertex $v$ a   \emph{level} $\ell_v$,
which serves as an exponentially smaller estimate to its out-degree, and to orient each edge towards the lower level endpoint.
It is instructive to view $\ell_v$ as      $\log (\dout(v))$, though in practice $\ell_v$ only serves as a ``proxy'' for $\log (\dout(v))$,
which intuitively means that it is being updated only after $\dout(v)$ has changed by a constant factor.
More accurately,
for any $\ell > \ell_v$, let $\phi_v(\ell)$ denote the number of neighbors of $v$ with level strictly lower than $\ell$.
At each update step, the levels of multiple vertices may change, so as to satisfy the following invariant for each vertex $v$:
\vspace{-2pt}
\begin{equation}
\forall \ell, \ell_v < \ell  \le \log (n-1): \phi_v(\ell) ~<~ 2^\ell .
\end{equation}
\vspace{-2pt}
If the invariant is violated, then $v$ will ``rise'' to the maximum level $\tilde \ell$ for which $\phi_v(\tilde \ell) \ge 2^{\tilde \ell}$.
As a result of this rise, incoming edges of $v$ from vertices at levels between $\ell_v$ and $\tilde \ell-1$ should be flipped out of $v$.

Since $v$'s outgoing neighbors are of level at most $\ell_v$, the invariant for $\ell = \ell_v + 1$ yields
$\dout(v) \le \phi_{v}(\ell_v + 1) < 2^{\ell_v+1}$, or equivalently $\ell_v > \log (\dout(v))-1$.
Symmetrically, to prevent $\ell_v$ from growing far beyond $\log (\dout(v))$,  vertices would   need to ``fall'' appropriately.  
However, any such vertex fall may trigger the rise of its neighbors, thus starting a long  cascade of vertices' levels rising and falling,
which may eventually result in many vertices  violating the invariant \emph{simultaneously}.
It is exactly for this reason that the invariant is stated for \emph{all} levels larger than $\ell_v$ rather than just for level $\ell_v+1$.
Indeed, \cite{BGS11} show that this much stronger form of the invariant gives rise to a \emph{fading wave mechanism}:
Following a fall of any vertex from level $\ell$ to level $\ell-1$, its neighbors may rise only to level $\ell$.
By breaking down the falling of a vertex into a series of one-level falls,
one can take care of vertices level by level, until the entire process ``fades away''. This beautiful observation lies at the core of Baswana et al.'s algorithm.


To maintain the invariant, one has to maintain the values $\phi_v(\ell)$, for all $\ell_v < \ell  \le \log (n-1)$.
Following a single edge update $(u,v)$, a logarithmic number of counters for $u$ and $v$ need to be updated,
requiring $O(\log n)$ time per update operation.
Even disregarding the maintenance time of these values, one    needs
to \emph{compute} the maximum level $\tilde \ell$ satisfying $\phi_v(\tilde \ell) \ge 2^{\tilde \ell}$.
Alas, due to the special nature of the $\phi_v(\cdot)$ values, this computation cannot be done via an efficient search method,
hence it requires a linear scan over a logarithmic number of optional levels.
Besides these  logarithmic-time bottlenecks, recall that the adversarial argument does not apply to
the induced deletions (by the algorithm) of matched edges.
Whenever a vertex $z$ is matched to a random neighbor $w$ that is already matched, say to vertex $w'$,
the cost of deleting edge $(w,w')$ from the matching can  be charged to the new matched edge $(z,w)$.
For this charging argument to work, it is crucial to have $\ell_w < \ell_z$, i.e.,
a mate should not be chosen randomly among all outgoing neighbors, but rather only among those with \emph{strictly lower} level.
The charging argument of \cite{BGS11} shows that the cost of induced deletions, when applied recursively, is logarithmic.
Indeed, as the ``fading wave'' may start at level $\Omega(\log n)$, it is only natural to incur a logarithmic time bound.

\vspace{4pt}
\noindent
{\bf Our approach.~} Our algorithm builds on the BGS scheme \cite{BGS11}, but is inherently different.
We too use a bucketing scheme with $O(\log n)$ levels, and orient each edge towards the lower level endpoint.  
However, our level-maintenance scheme is, in some sense, opposite to that of BGS. 
The invariant of \cite{BGS11} translates into $\log  (n-1) - \ell_v$ inequalities for each vertex $v$, which are maintained by the algorithm at all times.
By maintaining these inequalities, the level $\ell_v$ of each vertex $v$ is
kept close to $\log  (\dout(v))$, with a persistent attempt to rise to a higher level.
As opposed to \cite{BGS11}, we use a ``lazy'' approach, and in particular,  we do not try to satisfy any of these inequalities.
The level of a vertex $v$ will be changed by our algorithm only when absolutely necessary,
namely, following deletions or insertions of matched edges incident on $v$.

Consider a deletion of a matched edge $(z,w)$ at update step $l$,
suppose that $w$ was chosen uniformly at random among $z$'s outgoing neighbors as a mate for $z$ at some update step $l'$, for $l' < l$,
and denote $z$'s out-degree at step $l'$ by $\dout^{l'}(z)$.
Our algorithm sets $z$'s level $\ell_z$ at step $l'$ to   roughly $\log (\dout^{l'}(z))$, and leaves it unchanged throughout   update steps $l'+1,l'+2,\ldots,l$.
The expected number of outgoing edges of $z$ that are deleted from the graph during this time interval is $\dout^{l'}(z)/2$,
thus in the amortized sense, we should be able to cover an expected cost of $O(\dout^{l'}(z))$, for handling the deletion of edge $(z,w)$.

One consequence of our ``lazy'' approach is that we cannot bound the out-degree $\dout^{l}(z)$ of $z$ at step $l$ in terms of $\ell_z$ as in \cite{BGS11},
and in particular, it may be that $\dout^{l}(z) \gg 2^{\ell_z} \approx \dout^{l'}(z)$.
If it turns out that $\dout^{l}(z) = O(\dout^{l'}(z))$, then we can afford to scan all $z$'s outgoing neighbors.
Hence, we can handle $z$ \emph{deterministically}, i.e., match it with an arbitrary free outgoing neighbor, or leave it free if none exists.
(We will guarantee that no incoming neighbor of $z$ is free, hence we do not ignore free neighbors of $z$.)


The interesting case is when $\dout^{l}(z) \gg \dout^{l'}(z)$.
In this case we let $z$ rise to a higher level. However, we cannot let $z$ rise to the \emph{highest possible} level $\tilde \ell$ for which
$\phi_z(\tilde \ell) \ge 2^{\tilde \ell}$ as in \cite{BGS11}, since we do not maintain the values of $\phi_z(\cdot)$.
Instead, we   let $z$ rise \emph{gradually}, as long as $\phi_z(\ell) \ge 2^{\ell}$, stopping at the \emph{lowest} level $\ell^*$ after $\ell_z$ for which
$\phi_z(\ell^*+1) < 2^{\ell^* +1}$. If careless, the runtime of this rising process may be prohibitively large.
However, here we crucially exploit the fact that we have complete information of $z$'s incoming neighbors.
Indeed, using this information, we can restrict the attention to $z$'s neighbors of level at most $\ell^*$,
which enables us to implement the entire ``rising process'' in $\phi_z(\ell^*+1) = O(2^{\ell^*})$ time.
While this runtime may still be too large, we are guaranteed that the number of new outgoing neighbors of $z$
of level strictly lower than $\ell^*$ is $\phi_z(\ell^*) \ge 2^{\ell^*}$.
This means that we can cover the cost of the rising process
by  matching $z$ to a random such neighbor $w$.
Indeed, since the matched edge $(z,w)$ is chosen with probability at most $1/2^{\ell^*}$, we can cover an expected cost of  $\Omega(2^{\ell^*})$ in the amortized sense.
Summarizing:
\vspace{-3pt}
\begin{observation} \label{ob:key}
For every neighbor of $z$ scanned during this rising process, we spend $O(1)$ time.
To compensate for this, $z$'s mate is chosen uniformly at random among a constant fraction of these neighbors.
\end{observation}
\vspace{-3pt}

After matching $z$ with  $w$, we must let $w$ rise to the same level $\ell^*$ as $z$,
which requires flipping all incoming edges of $w$ from vertices at levels between $\ell_w$ and $\ell^* - 1$.
Alas, as another consequence of our lazy approach,
$w$'s  new out-degree may be much larger than $2^{\ell^*}$.
We can cover the cost of these flips by applying our level-rising mechanism on $w$, i.e., we let $w$ rise to yet a higher level,
stop at the lowest level $\ell'$ after $\ell^*$ for which $\phi_w(\ell'+1) < 2^{\ell'+1}$, and then match $w$ to a random outgoing neighbor of level   lower than $\ell'$.
To this end, however, we must first delete  edge $(z,w)$ from the matching, which requires handling $z$ (as a free vertex) from scratch.
Consequently, in contrast to the rather orderly manner in which vertices' levels are changed by the BGS algorithm
(particularly by the fading wave mechanism), the manner in which vertices' levels are changed by our algorithm (particularly by the level-rising mechanism) is rather chaotic.
Nevertheless, we show that our mechanism can be implemented via an elegant recursive algorithm, whose analysis boils down to a sophisticated application of Observation \ref{ob:key}.
In this way we bypass the use of the $\phi_v(\cdot)$ values,  ultimately achieving the optimal runtime and space bounds.  

As mentioned in Sect.\ 1.4, we believe that our level-rising mechanism will be applicable (after proper adjustments) to additional dynamic graph problems,
and it would be interesting to find such applications.

\vspace{-7pt}
\section{The Update Algorithm}
\vspace{-3pt}
The update algorithm is applied following edge insertions and deletions to and from the graph.
\vspace{7pt}
\\
\noindent
{\bf 3.1~ Invariants and data structures.~} Our algorithm maintains for each vertex $v$ a \emph{level} $\ell_v$, with $-1 \le \ell_v \le \log_3 (n-1)$.
(We use logarithms in base 3, whereas \cite{BGS11} use logarithms in base 2; this change simplifies the analysis, but is not crucial.)
The algorithm will maintain the following invariants.
\vspace{-2pt}
\begin{invariant} \label{first}
An edge $(u,v)$ with $\ell_u > \ell_v$ is \emph{oriented} by the algorithm as $\orient u v$. (In the case that $\ell_u = \ell_v$, the orientation of   $(u,v)$ will be determined
suitably by the algorithm.)
\end{invariant}

\vspace{-13pt}
\begin{invariant} \label{exception}
(a) Any free vertex has level -1 and out-degree 0. (Thus the maintained matching is maximal.)
~~(b) Any matched vertex has level at least 0.
~~(c) The two endpoints of any matched edge are of the same level, and this level remains unchanged until the edge is deleted from the matching.
(We may henceforth define the level of a matched edge, which is at least 0 by item (b), as the level of its endpoints.)
\end{invariant}
\vspace{-6pt}

For each vertex $v$, we maintain linked lists $\cN_v$ and $\cO_v$ of its neighbors and outgoing neighbors, respectively.  
The information about $v$'s incoming neighbors will be maintained via a more detailed data structure $\cI_v$:
A   hash table,  where each element corresponds to a distinct level $\ell \in \{-1,0,\ldots,\log_3 (n-1)\}$.
Specifically, an element $\cI_v[\ell]$ of $\cI_v$ corresponding to level $\ell$ holds a \emph{pointer} to the head of a non-empty linked list that
contains all incoming neighbors of $v$ with level $\ell$. 
If that list is empty, then the corresponding pointer is not stored in the hash table.
While the number of pointers stored in the  dynamic hash table $\cI_v$ is bounded by $\log_3 (n-1)+2$, it may be much smaller than that.
(Indeed, there is no pointer in $\cI_v$ corresponding to level smaller than $\ell_v$,
thus the number of pointers stored in $\cI_v$ is at most $\log_3 (n-1) + 1 -\ell_v$.
Also, by Invariant \ref{exception}, no incoming neighbor of $v$ is of level -1, thus there is no pointer in $\cI_v$ corresponding to level -1.)
In particular,   
the total space over all hash tables is linear in the dynamic number of edges
in the graph. 
We can use a static array of size $\log_3 (n-1) + 2$ instead of a dynamic  hash table, but   the total space usage over all these arrays will be $\Omega(n \log n)$.
In the case that the dynamic graph is usually dense (i.e., having $\Omega(n \log n)$ edges), it is advantageous to use arrays,
as it is easier to implement all the basic operations (delete, insert, search), and the time bounds become deterministic.

%
%
%
%

Note that no information whatsoever on the levels of $v$'s outgoing neighbors is provided by the data structure $\cO_v$.
In particular, to determine if $v$ has an outgoing neighbor at a certain level (most importantly at level -1, i.e., a free neighbor),
we need to scan the entire list $\cO_v$. On the other hand, $v$ has an incoming neighbor at a certain level $\ell$ iff
the corresponding list $\cI_v[\ell]$ is non-empty.
It will not be in $v$'s \emph{responsibility} to maintain the
data structure $\cI_v$, but rather within the responsibility of $v$'s incoming neighbors.

We keep mutual pointers between the elements in the various data structures: For any vertex $v$ and any outgoing neighbor $u$ of $v$,
we have mutual pointers between all elements $u \in \cO_v, v \in \cI_u[\ell_v],v \in N_u,u \in N_v$.
For example, when  an edge $(u,v)$ oriented as ${\orient v u}$ is deleted from the graph,
we get a pointer  to either $v \in N_u$ or $u \in N_v$, and through this pointer we delete all elements $u \in \cO_v, v \in \cI_u[\ell_v],v \in N_u,u \in N_v$.
As another example, when  the orientation of edge $(u,v)$ is flipped from ${\orient u v}$ to ${\orient v u}$, then assuming we have
a pointer to   $v \in \cO_u$ (which is  the case in our algorithm), we can reach element $u \in \cI_v[\ell_u]$ through the pointer, delete them both from the respective lists,
and then create elements $u \in \cO_v$ and $v \in \cI_u[\ell_v]$ with mutual pointers.
We also keep mutual pointers between a matched edge  and its endpoints.
(We do not provide a complete description of the trivial  maintenance of these pointers for the sake of brevity.)

Following \cite{BGS11}, we define $\phi_v(\ell)$ to be the number of neighbors of $v$ with level strictly lower than $\ell$.
\vspace{7pt}
\\
\noindent
{\bf 3.2~ Procedure $\setlvl(v,\ell)$.~}
Whenever the update algorithm examines a vertex $v$, it may need to re-evaluate its level.
After the new level $\ell$ is determined, the algorithm calls
Procedure $\setlvl(v,\ell)$. (See Figure \ref{fig:setlvl}.)
The procedure starts by updating the outgoing neighbors of $v$ about $v$'s new level.
Specifically, we scan the entire list $\cO_v$, and for each vertex $w \in \cO_v$, we move $v$ from $\cI_w[\ell_v]$ to $\cI_w[\ell]$.

Suppose first that $\ell < \ell_v$. In this case the level of $v$ is decreased by at least one.
As a result, we need to flip the outgoing edges of $v$ towards vertices of level between $\ell+1$ and $\ell_v$ to be incoming to $v$.
Specifically, we scan the entire list $\cO_v$,
and for each vertex $w \in \cO_v$ such that $\ell+1\le \ell_w \le \ell_v$, we
perform the following operations:
Delete $w$ from $\cO_v$, add $w$ to  $\cI_v[\ell_w]$, delete $v$ from $\cI_w[\ell]$, and add $v$ to $\cO_w$.

If $\ell > \ell_v$, the level of $v$ is increased  by at least one.
As a result, we
flip $v$'s incoming edges from vertices of level between $\ell_v$ and $\ell-1$ to be outgoing of $v$.
Specifically, for each non-empty list $\cI_v[i]$, with $\ell_v \le i \le \ell-1$,
and for each vertex $w \in \cI_v[i]$, we perform the following operations: Delete $w$ from $\cI_v[i]$, add $w$ to  $\cO_v$, delete $v$ from $\cO_w$, and add $v$ to $\cI_w[\ell]$.
Note, however, that we do not know for which levels $i$ the corresponding list is non-empty; the time overhead needed to verify this information is $O(\ell)$.

Only after the data structures have been updated, we set $\ell_v = \ell$.

The next observation is immediate from the description of the procedure, assuming Invariants \ref{first} and \ref{exception} hold.
In particular, it shows that the runtime of this procedure is at most $O(\dout^{old}(v) + \dout^{new}(v) + \ell)$, where
$\dout^{old}(v)$ and $\dout^{new}(v)$ denote $v$'s out-degree before and after the execution of this procedure, respectively.
\vspace{-3pt}
\begin{observation} \label{ob:setlevel}
Let $\ell_v$ denote the out-degree of $v$ before the execution of this procedure.
\begin{enumerate}
\item If $\ell = \ell_v$, the procedure does nothing, and the runtime is constant.
\item If $\ell < \ell_v$, then $\dout^{new}(v) \le \dout^{old}(v)$ and the procedure's runtime is $O(\dout^{old}(v))$.
\item If $\ell > \ell_v$,  then $\dout^{new}(v) \ge \dout^{old}(v)$ and the procedure's runtime is $O(\dout^{new}(v) + \ell)$.
Moreover,  after the execution of the  procedure, all outgoing neighbors of $v$ are of level at most $\ell-1$.
In the particular case of $\ell_v = -1$ and $\ell = 0$, which occurs when a free vertex $v$ becomes matched at level 0, 
we have $\dout^{new}(v) = 0$;
hence the procedure's runtime in this case is constant.
\end{enumerate}
\end{observation}
\vspace{1pt}
\noindent
{\bf 3.3~ Procedures $\ins(u,v)$ and $\del(u,v)$.~}
\\Following an edge insertion $(u,v)$, we apply Procedure $\ins(u,v)$; see Figure \ref{fig:handleIns} in App.\ \ref{app:tab}.
Besides updating the relevant data structures in the obvious way, this procedure matches between $u$ and $v$ if they are both free, or it leaves them unchanged.
Matching $u$ and $v$ involves setting their level to 0 by making the calls $\setlvl(u,0)$ and $\setlvl(v,0)$, whose runtime is $O(1)$ by Observation \ref{ob:setlevel}(3).

Following an edge deletion $(u,v)$, we apply Procedure $\del(u,v)$; see Figure \ref{fig:handleDel}.
If edge $(u,v)$ does not belong to the matching,   we only need to update the relevant data structures.
In the case  that  edge $(u,v)$ belongs to the matching, both $u$ and $v$ become \emph{temporarily free},
meaning that they are not matched to any vertex yet, but their level remains temporarily as before.
(We   handle them next, one after another, but until each of them is handled, its level will exceed -1.)
We handle vertices $u$ and $v$ via Procedure $\evalf$, specifically, by calling $\evalf(u)$ and later $\evalf(v)$;
Procedure $\evalf$ is the main ingredient of the update algorithm, and is described in Section 3.4.
\vspace{7pt}
\\
\noindent
{\bf 3.4~ Procedure $\evalf(v)$.~}
The execution of this procedure splits into two cases. See Figure \ref{fig:evalfree}.
\\\emph{Case 1:} $\dout(v)  < 3^{\ell_v+1}$.
In other words, the first case is when the out-degree of $v$ is not much greater than $3^{\ell_v}$,
and we run Procedure $\detr(v)$ described in Section
3.4.1 (see Figure \ref{fig:detr}).
\\\emph{Case 2:} $\dout(v) \ge 3^{\ell_v+1}$.
We run Procedure $\rand(v)$ described in Section
3.4.2 (see Figure \ref{fig:rand}).
\vspace{-8pt}
\\
\noindent
{\bf 3.4.1~ Procedure $\detr(v)$.~}
The procedure starts by scanning the list $\cO_v$ for a free vertex.
By Invariant \ref{exception}(a), if no free vertex is found in $\cO_v$,  then $v$ does not have any free neighbor.
(Note that we ignore temporarily free neighbors of $v$. See Sect.\ 4.1 for more details.)
If no free vertex is found in $\cO_v$, then $v$ becomes free and we set its level $\ell_v$ to -1 by calling to $\setlvl(v,-1)$.

Otherwise a free vertex is found  in $\cO_v$. In this case we match $v$ to an arbitrary such vertex $w$, and set the levels of $v$ and $w$ to 0
by calling to $\setlvl(v,0)$ and $\setlvl(w,0)$. 

Next, we show that the runtime of the procedure is bounded by $O(3^{\ell_v})$. 
Let $\dout^{old}(v)$ denote $v$'s out-degree before the execution of the procedure, 
and note that this procedure is invoked only when $\dout^{old}(v)  < 3^{\ell_v+1}$.
Moreover, since $v$ is a temporarily free vertex, its level at this stage is at least 0.

The procedure starts by scanning the list $\cO_v$ for a free vertex, which takes $O(\dout^{old}(v)) = O(3^{\ell_v})$ time.
Next, the procedure calls to either  $\setlvl(v,-1)$ or $\setlvl(v,0)$.
Since the level of $v$ prior to either one of these calls is at least 0,
$v$'s level may only decrease, hence the runtime is bounded by $O(\dout^{old}(v)) = O(3^{\ell_v})$ by Observations \ref{ob:setlevel}(1) and \ref{ob:setlevel}(2).
Finally, there is   a potential call to $\setlvl(w,0)$, which increases the level of $w$ from -1 to 0; the runtime of this call is constant by Observation \ref{ob:setlevel}(3).

We remark that the out-degree of $v$ after the potential call to $\setlvl(v,0)$ may be large,
since it may have many outgoing neighbors of level 0. However, by Observations \ref{ob:setlevel}(1) and \ref{ob:setlevel}(2), this out-degree is bounded by $\dout^{old}(v)  < 3^{\ell_v+1}$.
Even though we can afford to flip all the edges leading to those vertices (this would require at most $O(3^{\ell_v})$ time, which we spend anyway), there is no need for it.
\vspace{6pt}
\\
\noindent
{\bf 3.4.2~ Procedure $\rand(v)$.~} The procedure employs what we refer to as a \emph{level-rising mechanism}.
Roughly speaking, the procedure matches $v$ at some level $\ell^*$ higher than $\ell_v$, with a random (possibly matched) neighbor of level \emph{strictly lower} than $\ell^*$.
More accurately, it \emph{attempts} to create such a matched edge $(v,w)$ at level $\ell^*$ within time $O(3^{\ell^*})$;
upon failure, it calls itself recursively to match $w$ at yet a higher level, in which case $v$ becomes free and handled via Procedure $\evalf(v)$.

Next, we describe this procedure, and the underlying level-rising mechanism,  in detail.
We find it instructive to provide this description in two stages. 
In the first stage we outline the two main challenges that this procedure has to cope with,
and the specific manner in which it copes with these challenges.  
Only after the appropriate intuition is established, we turn to the formal description of the procedure.
\vspace{7pt}
\\
\noindent
{\bf Challenge 1.~}
Recall that $\phi_{v}(\ell)$ is the number of $v$'s neighbors of level \emph{strictly lower} than $\ell$.
The reason we restrict our attention to neighbors of $v$ of level \emph{strictly lower} than a certain threshold is fundamental.
When choosing a  random mate $w$ for $v$, we cannot guarantee that $w$ would be free.
Assuming $w$ is matched to $w'$, matching $v$ with $w$ triggers the deletion of edge $(w,w')$ from the matching.
This edge deletion is \emph{induced} by the algorithm itself rather than the adversary, i.e., the edge remains in the graph but deleted from the matching.
Alas, the adversarial argument, which bounds the expected number of edges that are deleted from the graph until the matched edge is deleted from the matching,
does not hold for induced deletions.
Coping with induced deletions, which is the crux of the problem, requires an intricate charging argument.
As in \cite{BGS11}, when choosing a mate $w$ for $v$, we restrict our attention 
to $v$'s neighbors of level strictly lower than that of $v$, 
or more accurately, strictly lower than the \emph{new} level to which $v$ rises following the random match.
This restriction, however, poses a nontrivial challenge.

One cannot simply set the new level of $v$ to be sufficiently large.
Indeed, for our probabilistic argument to work, it is crucial that $v$'s new level would depend on the
probability with which the new matched edge is chosen: If $v$'s new level is $\ell^*$,
the probability that $w$ is chosen as $v$'s mate should be $O(1/3^{\ell^*})$.
To guarantee that this condition holds, though, $v$ must have $\Omega(3^{\ell^*})$ neighbors of level strictly lower than $\ell^*$.

It is   easy to show that such a level $\ell^*$ exists (see Lemma \ref{rand_basic}(1)).
It is much less clear, however, how to compute it efficiently.
The challenge that we face is actually more complex:
After setting the level $\ell_v$ of $v$ to $\ell^*$, we need to update the data structures accordingly.
For this scheme to work, the entire runtime should be $O(3^{\ell^*})$.
To see where the difficulty lies, suppose we take $\ell^*$ to be $\ell_v$.
Recalling that $\dout(v) \ge 3^{\ell_v+1}$, $v$ has   many neighbors of level at most $\ell^*$.
However, we need that $v$ would have many neighbors of level \emph{strictly lower} than $\ell^*$, and it is possible that most (or all)
neighbors of $v$ have level $\ell^*$. One may try taking $\ell^*$ to be $\ell_v + 1$.
This would indeed guarantee that $v$ has sufficiently many neighbors of level strictly lower than $\ell^* + 1$.
However, now there in another, somewhat contradictory, problem.
After setting $\ell^*$ to $\ell_v + 1$, we need to update the data structures.
In particular, we must flip all the incoming edges of $v$ from neighbors of level $\ell_v$ to be outgoing of $v$.
This may be prohibitively expensive.

In general, the ``right'' level $\ell^*$ should balance two contradictory requirements:
While $v$ should have sufficiently many neighbors of level strictly lower than $\ell^*$, 
it should not have too many of them.
Any level balancing these requirements will do the job, but we also need to be able to compute it efficiently.
We next show that, somewhat surprisingly, a sequential scan for the right level works smoothly.
\vspace{4pt}
\\
\noindent
{\em Computing the ``right'' level for $v$:~~}
Set $\ell = \ell_v$, and gradually increase $\ell$ as long as $\phi_{v}(\ell) \ge 3^{\ell}$.
Let $\ell^*$ be the level in which we stop the process, i.e.,  the \emph{minimum} level such that $\ell^* \ge \ell_v$ and  $\phi_v(\ell^*+1) < 3^{\ell^*+1}$.
We then set $v$'s level  to $\ell^*$ by calling $\setlvl(v,\ell^*)$, thus updating the data structures,
which   involves flipping all   incoming edges of $v$ from neighbors of level between $\ell_v$ and $\ell^* - 1$ to be outgoing of $v$.

The following lemma is crucial to the correctness of the level-rising mechanism.  
It shows that the level $\ell^*$ satisfies several somewhat contradictory requirements, each of which is important for our algorithm.
We stress that, to be able to efficiently compute the level $\ell^*$, we make critical use of the fact that we have complete information
of the incoming neighbors of $v$; see the proof of Lemma \ref{rand_basic} for more details.
\vspace{-2pt}
\begin{lemma} \label{rand_basic}
Let $\dout^{old}(v)$ and $\dout^{new}(v)$ denote $v$'s out-degree before and after the call to  $\setlvl(v,\ell^*)$, respectively.
Here $\ell_v$ denotes the level of $v$ before the call, whereas $\ell^*$ is its level afterwards.
\vspace{4pt}\\(1) $\ell_v < \ell^* \le \log_3 (n-1)$. In particular, a level $\ell^*$ as required exists.
Moreover, the call to $\setlvl(v,\ell^*)$ increases $v$'s level by at least one.
\vspace{4pt}\\(2)  After this call, all outgoing neighbors of $v$ are of level at most $\ell^*-1$, i.e., $\phi_v(\ell^*) = \dout^{new}(v)$.
\vspace{4pt}\\(3) $\dout^{old}(v) \le \dout^{new}(v)$ and $3^{\ell^*} \le  \dout^{new}(v) = \phi_v(\ell^*) \le \phi_v(\ell^*+1) < 3^{\ell^* + 1}$. (In particular, $\dout^{old}(v) < 3^{\ell^* + 1}$.)
\vspace{-6pt}\\(4) The runtime of computing  $\ell^*$ and calling to  $\setlvl(v,\ell^*)$ is bounded by $O(3^{\ell^*})$.
\end{lemma}
\vspace{-2pt}
\begin{proof}
(1)  By invariant 1, all outgoing neighbors of $v$ before the call have out-degree at most $\ell_v$,
so $\phi_v(\ell_v+1) \ge \dout^{old}(v)$. We also have $\dout^{old}(v) \ge 3^{\ell_v+1}$,
yielding  $\phi_v(\ell_v+1) \ge \dout^{old}(v) \ge 3^{\ell_v+1}$, and so $\ell^* > \ell_v$.
Since $\phi_v(\log_3 (n-1)+1) \le \deg(v) \le n-1 < 3^{\log_3 (n-1) + 1}$, it follows that $\ell^* \le \log_3 (n-1)$.
\vspace{4pt}\\(2) By the first assertion of this lemma, $\ell^* > \ell_v$.
This assertion thus follows from Observation \ref{ob:setlevel}(3).
\vspace{4pt}\\(3) The first assertion of this lemma and Observation \ref{ob:setlevel}(3) yield $\dout^{old}(v) \le \dout^{new}(v)$.
The second assertion of this lemma  and the definitions of $\ell^*$ and $\phi_v(\cdot)$ yield  $3^{\ell^*} \le  \dout^{new}(v) = \phi_v(\ell^*) \le \phi_v(\ell^*+1) < 3^{\ell^* + 1}$.
\vspace{4pt}\\(4) Recall that we have complete information of the incoming neighbors of $v$ via the data structure $\cI_v$.
This enables us to restrict our attention to the neighbors of $v$ up to a certain level, and ignore the others.
Specifically, for each non-empty list $\cI_v[\ell]$, with $\ell = \ell_v, \ell_v+1, \ldots$, we naively count the number of vertices in $\cI_v[\ell]$
in order to construct the values $\phi_v(\ell)$, for $\ell = \ell_v, \ell_v+1, \ldots$. Stopping this process once reaching $\ell^*+1$,
the time spent is at most linear in the number of scanned levels plus the number of vertices in the corresponding lists $\cI_v[\ell]$.
The number of scanned levels is bounded by $\ell^*+1$ and the number of traversed vertices is bounded by $\phi_v(\ell^*+1)$, which is smaller than $3^{\ell^* + 1}$ by the
third assertion of this lemma. 

By being able to restrict the attention to $v$'s neighbors of level at most $\ell^*+1$,
we have shown that the runtime of computing the ``right'' level $\ell^*$ is   $O(3^{\ell^*})$.
Moreover, by Observation \ref{ob:setlevel}(3), the runtime of the call to $\setlvl(v,\ell^*)$ is bounded by $O(\dout^{new}(v) + \ell^*)$, which is, in turn,
at most $O(3^{\ell^*})$ by the third assertion of this lemma. 
We stress that the validity of Observation \ref{ob:setlevel}(3) is also based on our ability to restrict the attention to the (incoming) neighbors of $v$
whose level is bounded by some threshold.  
\QED
\end{proof}
\vspace{4pt}
\noindent
{\bf Challenge 2.~}
Lemma \ref{rand_basic} implies that $O(3^{\ell^*})$ time suffices for
computing the ``right'' level $\ell^*$ and letting $v$ rise to that level by making the call $\setlvl(v,\ell^*)$.
Moreover, $v$ has at least $3^{\ell^*}$ outgoing neighbors  after this call, all having level at most $\ell^* - 1$.
By picking uniformly at random an outgoing neighbor $w$ of $v$ to match with, we are guaranteed that the matched edge $(v,w)$ is chosen with probability at most $1/3^{\ell^*}$.
In the amortized sense, this matched edge can cover the entire cost $O(3^{\ell^*})$ spent thus far.

In order to add edge $(v,w)$ to the matching, however, we need to update the  data structures accordingly.
In particular, if $w$ is matched, say to $w'$,   we must delete edge $(w,w')$ from the matching; this is an \emph{induced} edge deletion.
As mentioned, 
to cope with induced deletions,
our charging argument makes critical use of the fact that $w$ is of level strictly lower than $\ell^*$.
However, this requirement by itself would suffice  only if we were guaranteed that edge $(w,w')$ was chosen to the matching by $w$.
(If the vertex initiating the match  is of level $\ell$, then the cost of creating the matched edge is $O(3^{\ell})$ by Lemma \ref{rand_basic}.)
In general, this edge might have been chosen to the matching by $w'$ rather than $w$,
and so it is critical that both endpoints $w$ and $w'$ would be of levels strictly lower than $\ell^*$ for the charging argument to work.
To this end, as in \cite{BGS11}, we maintain the stronger invariant that the two endpoints of any matched edge are of the same level; see Invariant \ref{exception}(c).
Consequently, to match $v$ with $w$, the invariant requires that we let $w$ rise to the new level $\ell^*$ of $v$. 
This requirement, however, poses another nontrivial challenge.

We set the level $\ell_w$ of $w$ to $\ell^*$ by calling $\setlvl(w,\ell^*)$.
This call updates the data structures accordingly, which  involves flipping all   incoming edges of $w$ from neighbors of level between $\ell_w$ and $\ell^* - 1$ to be outgoing of $w$.
As before, this may be prohibitively expensive.
Specifically, by Observation \ref{ob:setlevel}(3), the runtime of the call to $\setlvl(w,\ell^*)$ is $O(\dout(w) + \ell^*)$, where $\dout(w)$ is the new out-degree of $w$.

By setting the level of $w$ to $\ell^*$, we have created a matched edge $(v,w)$ at level $\ell^*$. 
If and when this matched edge is  deleted from the graph, we will be able to cover an expected cost of $O(3^{\ell^*})$ in the amortized sense.
If $\dout(w) < 3^{\ell^*+1}$, the runtime of   the call to $\setlvl(w,\ell^*)$,
and thus of the entire procedure, is $O(3^{\ell^*})$, which can be covered in the amortized sense by the creation of the new matched edge $(v,w)$.
However, the complementary case $\dout(w) \ge  3^{\ell^*+1}$ is where the   difficulty lies.
Indeed, in this case the runtime of the call to $\setlvl(w,\ell^*)$
may be significantly higher than $3^{\ell^*}$.
To cover it, we  create a matched edge at level higher than $\ell^*$.
To this end we first delete the new matched edge $(v,w)$ from the matching,
and then invoke Procedure $\rand$ recursively, but on $w$ this time.

The recursive call $\rand(w)$ creates a matched edge at level higher than $\ell^*$, which, in the amortized sense, can cover the cost of the call to $\setlvl(w,\ell^*)$.
Thus, in each recursive call we rise to yet a higher level, attempting to charge the yet-uncharged costs of the procedure to the most recently created matched edge.
We stress that the level-rising mechanism is not a single computation of a matched edge at some ``right'' level,
but rather a recursive attempt at doing so: Try, rise to a higher level upon failure, and then try again.
Note that the maximum level is $\log_3 (n-1)$. Since $\dout(w) \le \deg(w) \le n-1 < 3^{\log_3 (n-1)+1}$, for any vertex $w$, this recursive attempt eventually succeeds.
\vspace{7pt}
\\
\noindent 
{\bf The procedure.}
Procedure $\rand(v)$ starts by computing the level $\ell^*$ as described above
(i.e., the first level after $\ell_v$ such that $\phi_v(\ell^*+1) < 3^{\ell^*+1}$)
and setting $v$'s level accordingly by calling $\setlvl(v,\ell^*)$.
We then pick uniformly at random an outgoing neighbor $w$ of $v$, to match them.
(Lemma \ref{rand_basic}(3) implies that the matched edge is chosen with probability at most $1/3^{\ell^*}$,
whereas $\ell_w \le \ell^*-1$ follows from Lemma \ref{rand_basic}(2).)
If $w$ is matched, say to $w'$, then we delete edge $(w,w')$ from the matching.
This renders $w'$ \emph{temporarily free}, meaning that it is not matched to any vertex, but its level remains temporarily as before;
$w'$ will be handled soon, but  in the interim, its level will exceed -1. (See Figure \ref{fig:rand}.)

We set the level of $w$ to $\ell^*$ by calling $\setlvl(w,\ell^*)$, which increases its level by at least one,
and add edge $(v,w)$ to the matching,
thus creating a matched edge of level $\ell^*$. 
(If and when this matched edge is  deleted from the graph, we will be able to cover an expected cost of $O(3^{\ell^*})$ in the amortized sense.)

The runtime of the call to $\setlvl(w,\ell^*)$ is $O(\dout(w) + \ell^*)$, where $\dout(w)$ is the new out-degree of $w$.
If $\dout(w) < 3^{\ell^*+1}$, this runtime is $O(3^{\ell^*})$, and it can be covered in the amortized sense.

However, in the complementary case $\dout(w) \ge  3^{\ell^*+1}$, the runtime
may be significantly higher than $3^{\ell^*}$.
To cover this runtime, we  create a matched edge at level higher than $\ell^*$.
To this end we first delete the new matched edge $(v,w)$ from the matching,
thus rendering $v$ and $w$ temporarily free, and then invoke   Procedure $\rand$ recursively, by calling $\rand(w)$.
(This recursive call creates a matched edge at level higher than $\ell^*$, which, in the amortized sense, can cover the cost of the call to $\setlvl(w,\ell^*)$.)
It is possible that $v$ will become matched as a result of the recursive call to $\rand(w)$;
if $v$ is not matched to any vertex, we invoke Procedure $\evalf(v)$.

Finally, if $w'$ is not matched to any vertex, we invoke Procedure $\evalf(w')$.

\vspace{-6pt}
\section{Analysis} \label{sec:mainanal}
\vspace{-3pt}
{\bf 4.1~ Invariants.~}
It is easy to verify that our update algorithm satisfies Invariants \ref{first} and \ref{exception}.
The only (technical) exception to Invariant \ref{exception} is with temporarily free vertices, which are unmatched, yet their level exceeds -1.
A vertex becomes temporarily free after its matched edge is deleted, either by the adversary (see line 5(a) in Figure \ref{fig:handleDel}) or via Procedure $\rand$
of the update algorithm (see lines 6 and 9(a) in Figure \ref{fig:rand}).
When a free vertex $v$ is handled via Procedure $\evalf(v)$, it may \emph{ignore} its temporarily free neighbors, as the corresponding edges may
be incoming to $v$. 
In particular, Procedure $\detr(v)$ \emph{deliberately ignores} the free neighbors of $v$, and as a result,
$v$ may become free although it may have temporarily free neighbors.
(Obviously, this is a matter of choice; we can change the procedure to consider temporarily free neighbors of $v$ that belong to $\cO_v$, but there is no need.)
For this reason, our update algorithm makes sure to handle all vertices that become temporarily free \emph{later}, via appropriate calls to Procedure $\evalf$; see lines 5(b) and 5(c) in Figure \ref{fig:handleDel} and lines 9(c) and 10 in Figure \ref{fig:rand}.
Hence, if any  temporarily free neighbor $w$ of $v$ is ignored by $v$ and $v$ is left free,
the subsequent call to $\evalf(w)$ will match $w$, either with $v$ or with another    neighbor of $w$.
\vspace{7pt}
\\
\noindent
{\bf 4.2~ Epochs.~}
Given any sequence of edge updates, an edge $(u,v)$ may become matched or unmatched by the algorithm at different update steps.
The entire lifespan of an edge $(u,v)$ consists of a sequence of \emph{epochs}, which refer to the maximal time intervals in which the edge is matched,
separated by
the maximal time intervals in which the edge is unmatched.
(The notion of an epoch was introduced in \cite{BGS11}.)  
Formally, let $e = (u,v)$ be any edge of $\cM = \cM_l$ at some time step $l$.
The \emph{epoch} $\cE(e,l)$ corresponding to edge $e$ at time $l$ refers to the maximal time interval containing $l$
during which $(u,v) \in \cM$.

An epoch $\cE(e,l)$ is not just a time interval, but rather an object describing a \emph{specific edge} within that time interval.
In particular, for any two distinct (matched) edges $e$ and $e'$ and any time $l$, the respective epochs $\cE(e,l)$ and $\cE(e',l)$ refer to different objects.
On the other hand, for two distinct times $l$ and $l'$ and some edge $e$, it is possible that the respective epochs $\cE(e,l)$ and $\cE(e,l')$
refer to the same object.

By Invariant \ref{exception}(c),  the endpoints of a matched edge are of the same level, and this level remains unchanged.  
We   henceforth define the level of an epoch to be the level of the corresponding edge.


Any edge update that does not change the matching is processed by our algorithm in constant time.
However, an edge update that changes the matching may trigger the creation of some epochs and the termination of some other epochs.
The computation cost of creating or terminating an epoch by the algorithm may be large. Moreover, the number of epochs created and terminated due to
a single edge update may be large by itself. Therefore, an amortized analysis is required.
Following the amortization scheme of \cite{BGS11},
we re-distribute the total computation performed at any  step $l$ among the epochs created or terminated at step $l$.
Specifically, let $\cE_1 = \cE(e_1,l),\ldots,\cE_j = \cE(e_j,l)$ (resp., $\cE'_1 = (e'_1,l),\ldots,\cE'_k = (e'_k,l)$) be the epochs created (resp., terminated) at update step $l$, and let $C_{create}(\cE_1),\ldots,C_{create}(\cE_j)$ (resp., $C_{term}(\cE'_1),\ldots,C_{term}(\cE'_k)$) be the respective
computation costs charged to the creation (resp., termination) of these epochs.
Then we re-distribute the total computation cost performed at update step $l$, denoted by $C^{(l)}$,
to the respective epochs in such a way that $C^{(l)} = \sum_{i=1}^j C_{create}(\cE_i) + \sum_{i=1}^k C_{term}(\cE'_i)$.

\vspace{-6pt}
\begin{claim} \label{levelepoch}
(1) Any epoch created by Procedures $\ins$ or $\detr$ is of level 0.
~~(2) Any epoch created by Procedure $\rand$ is of level at least 1.
\end{claim}
\vspace{-6pt}
\begin{proof}
(1) Immediate.
~~(2) Consider a vertex $v$ that is handled via Procedure $\rand(v)$. First note that $\dout(v) \ge 3^{\ell_v+1} \ge 1$.
By Invariant \ref{exception}, $\ell_v \ge 0$.
By the description of Procedure $\rand(v)$ and Lemma \ref{rand_basic}(1),
we conclude that any epoch created by this procedure is of level at least $\ell^* > \ell_v$.
\QED
\end{proof}
Lemma \ref{rand_basic} and Claim \ref{levelepoch} yield the following corollary.
\vspace{-3pt}
\begin{corollary} \label{Ob:basic}
For any epoch at level $\ell > 0$, initiated by vertex $v$ and corresponding to edge $(v,w)$:
\\(1) The out-degree of $v$ at the time the epoch is created is at least $3^{\ell}$ and less than $3^{\ell+1}$,
though the out-degree of $w$ at that time
may be significantly smaller or larger than $3^{\ell}$.
\\(2) $w$ is chosen as a mate for $v$ uniformly at random among all outgoing neighbors of $v$ at that time,
hence the corresponding edge $(v,w)$ becomes matched with probability at most $1/3^{\ell}$.  
\end{corollary}

\vspace{-0pt}
\noindent{\bf 4.3~ Re-distributing the computation costs to epochs.~}
By re-distributing the total computation cost of the update algorithm to the various epochs, we can \emph{visualize} the entire update algorithm as a sequence of creation and termination of these epochs.
The computation cost associated with an epoch at level $\ell$ (hereafter,  \emph{level-$\ell$ epoch}) includes both its creation cost and its termination cost.

The following lemma plays plays a central role in our analysis. 
Although a similar lemma was proved in \cite{BGS11}, our proof is inherently different than the corresponding proof of \cite{BGS11}.
\vspace{-1pt}
\begin{lemma} \label{basiccost}
The total computation cost of the update algorithm 
can be re-distributed to various epochs
so that the computation cost 
associated with any {level-$\ell$ epoch}
 is bounded by $O(3^{\ell})$, for any $\ell \ge 0$.
\end{lemma}
\vspace{-3pt}
\begin{proof}
The update algorithm is triggered following edge insertions and edge deletions.

Following an edge insertion, we apply
Procedure $\ins(u,v)$. This procedure occurs at most once per update step, and its runtime is constant.
We may disregard this constant cost (formally, we charge this cost to the corresponding update step).
Hence, the lemma holds vacuously for edge insertions.

Following an edge deletion, we apply
Procedure $\del(u,v)$. This procedure occurs at most once per update step; disregarding the calls to $\evalf(u)$ and $\evalf(v)$, the runtime of this procedure is constant as well.
We henceforth disregard this constant cost (charging it to the corresponding update step), and demonstrate how to re-distribute the costs of the calls to $\evalf(u)$ and $\evalf(v)$ to 
appropriate epochs in a manner satisfying the condition of the lemma.

Let $z \in \{u,v\}$. The execution of Procedure $\evalf(z)$ splits into two cases.
In the first case $\dout(z) < 3^{\ell_z+1}$, and Procedure $\evalf(z)$ invokes Procedure $\detr(z)$, whose runtime is $O(3^{\ell_z})$.
Even though Procedure $\detr(z)$ may create a new level-0 epoch, there is no need to charge this epoch with any costs.
The entire cost $O(3^{\ell_z})$ of Procedure $\detr(z)$ is charged to the termination cost of epoch $\cE((u,v),l)$ (triggered by the deletion of edge $(u,v)$ from the graph).
 which is of level $\ell_z$, thus satisfying the condition of the lemma.

Otherwise $\dout(z) \ge 3^{\ell_z+1}$, and Procedure $\evalf(z)$ invokes Procedure $\rand(z)$.

The runtime of Procedure $\rand(z)$ may be much larger than $3^{\ell_z}$ and even than $\dout(z)$.
To cover the costs of this procedure, we create a new epoch at a sufficiently high level, attempting to charge the costs of the procedure to the creation cost of the new epoch.
However, this charging attempt may sometimes fail, in which case we call the procedure recursively.
Each recursive call will create a new epoch at yet a higher level,  attempting to charge the yet-uncharged costs of
the procedure to the creation cost of the most recently created epoch.
Since the levels of epochs created during this process grow by at least one  with each recursion level, this recursive process will terminate.
Specifically, the depth of this recursive process is bounded by $\log_3 (n-1)$.
(Formally, one should also take into account the calls to Procedure $\evalf$ from within Procedure $\rand$, which may, in turn, invoke Procedure $\rand$.\footnote{Since the assertion
that the recursive process terminates is a corollary of our ultimate bound on the update time,
an additional proof is not required. Nevertheless, we next sketch a simpler proof
that is  independent of our update time bound.
We define a  \emph{potential function} $f(G) = \sum_{v \in V} {3^{2\ell_v}}$ for the dynamic graph $G$ with respect to the dynamic level assignment of its vertices.
The initial call to Procedure $\rand$ may trigger a single call to Procedure $\evalf$.
Each subsequent recursive call to Procedure $\rand$ may trigger two calls to Procedure $\evalf$.
It can be easily verified that each call to Procedure $\rand$ must increase the potential by at lease one unit more than the calls to Procedure $\evalf$ that it triggers may decrease it.
Each such call to   $\evalf$ may, in turn, invoke Procedure $\rand$. 
Nevertheless, the potential growth due to each such call to Procedure $\rand$ is at least 1,
regardless of whether it is a recursive call invoked by Procedure $\rand$ itself
or a new call invoked by Procedure $\evalf$.
Since the potential value is upper bounded by $n \cdot 3^{2\log_3(n-1)} \le n^{3}$ at all times, it follows that the total number of calls to Procedure $\rand$
is   upper bounded by $n^3$; in particular, this number is finite, thus the recursive process must terminate.})
Next, we describe the charging argument in detail.

For each recursion level of Procedure $\rand$,
our charging argument may (slightly) \emph{over-charge} the current level's costs,
so as to cover some of the yet-uncharged costs incurred by  previous recursion levels.
Denote by $\rand(z^{(i)})$ the $i$th recursive call, with $z^{(i)}$ being the examined vertex.
The initial call to Procedure $\rand$ may be viewed as the 0th recursion level.  
As we show below, it may leave a ``debt''
to the 1st recursion level, but this debt must be bounded by $O(\dout(z^{(1)}))$ units of cost.
In general, for each $i \ge 1$, the \emph{debt} at recursion level $i$ (left by previous recursion levels) must be bounded by  $O(\dout(z^{(i)}))$ units of cost;
in what follows we may view this debt as part of the costs incurred at level $i$. (Note that the initial call to the procedure has no debt.)

Observe that the $i$th recursive call $\rand(z^{(i)})$ may also invoke Procedure $\evalf$.
Our charging argument views the call to   $\evalf$ as part of the execution of $\rand(z^{(i)})$.
More concretely, the costs incurred by the call to $\evalf$ are charged to epochs that are created or terminated throughout the execution of the $i$th recursive call $\rand(z^{(i)})$.
Note, however, that Procedure $\evalf$ may, in turn, invoke Procedure $\rand$. 
Our charging argument does not view the new call to $\rand$ as part of the execution of $\rand(z^{(i)})$, but rather as an \emph{independent} call.
More concretely, the costs of the new call to Procedure $\rand$ are 
charged only to epochs that are created or terminated throughout the execution of the new call to   $\rand$.


The following claim completes the charging argument, thus concluding the proof of Lemma \ref{basiccost}.
\begin{claim} \label{conc}
For any $i \ge 0$, we can re-distribute the costs of the $i$th recursive call $\rand(z^{(i)})$ of Procedure $\rand$ (including the debt from previous recursion levels, if any)
to various epoch, allowing a potential debt to the subsequent recursive call $\rand(z^{(i+1)})$ (if the procedure proceeds to recursion level $i+1$),
so that (1) The cost charged to any level-$\ell$ epoch is bounded by $O(3^{\ell})$, for any $\ell$, and
(2) The debt left for recursion level $i+1$ is bounded by $O(\dout(z^{(i+1)}))$. 
If the procedure terminates at recursion level $i$, then no debt is allowed, i.e., the entire cost in this case is re-distributed to the epochs.
\end{claim}
\begin{proof}
The proof is by induction on the recursion level $i$.
Recall that the initial call $\rand(z^{(0)})$ of the procedure, which corresponds to the basis $i=0$ of the induction, has no debt.
Nevertheless, the basis of the induction and the induction step are handled together as follows.

Consider the $i$th recursive call $\rand(z)$, $i \ge 0$, writing $z = z^{(i)}$ to avoid cluttered notation.
Observe that the debt from previous recursion levels is bounded by $O(\dout(z))$ units of cost. 
(This observation follows from the induction hypothesis for $i-1$, unless $i=0$, in which case it holds vacuously.)
To cover this debt, the procedure creates a new epoch of sufficiently high level,
determined as the minimum level $\ell^* = {\ell^*}^{(i)}$ after $\ell_z$ such that $\phi_z(\ell^*+1) < 3^{\ell^*+1}$.
By Lemma \ref{rand_basic}(4), computing this level $\ell^*$ requires $O(3^{\ell^*})$ time, which, by Lemma \ref{rand_basic}(3),
already supersedes the debt of $O(\dout(z))$ units of cost from previous recursion levels; we may henceforth disregard this debt.
The procedure then sets the level of $z$ to $\ell^*$ by calling to $\setlvl(z,\ell^*)$.
By Lemma \ref{rand_basic}(4), the runtime of this call is $O(3^{\ell^*})$.
Then a random outgoing neighbor $w$ of $z$  is chosen as a mate for $z$,
and its level is set to $\ell^*$ by calling to $\setlvl(w,\ell^*)$.
By Observation \ref{ob:setlevel}(3), the runtime of this call is $O(\dout(w) + \ell^*)$, where $\dout(w)$ is $w$'s new out-degree, and it may be that $\dout(w) \gg 3^{\ell^*}$.
In the case that $\dout(w) < 3^{\ell^*+1}$, disregarding a potential call to $\evalf(w')$ that is addressed below,
the entire cost of the $i$th recursive call  $\rand(z^{(i)})$  of the procedure is $O(3^{\ell^*})$, and we charge it to the creation cost of the new  level-$\ell^*$ epoch $\cE((z,w),l)$.
In this case the procedure terminates at recursion level $i$, and no debt whatsoever it left.

If $\dout(w) \ge 3^{\ell^*+1}$, Procedure $\rand(z)$ deletes the new matched edge $(z,w)$ from the matching, thus terminating the respective epoch
$\cE((z,w),l)$.
Then it proceeds to recursion level $i+1$  by making a recursive call to $\rand(w)$,  with $w = z^{(i+1)}$.
By Lemma \ref{rand_basic}(1), this recursive call is guaranteed to create a new epoch at level $\ell'$ higher than $\ell^*$.
Moreover, Lemma \ref{rand_basic}(3) implies that $\dout(w) < 3^{\ell' + 1}$,
and so the cost $O(\dout(w) + \ell^*) = O(\dout(w)) = O(3^{\ell'})$ of the call to $\setlvl(w,\ell^*)$ made in the $i$th recursion level can be charged to the creation cost of the  new level-$\ell'$ epoch
that is created in recursion level $i+1$.
Formally, we ``drag'' the cost of the call to $\setlvl(w,\ell^*)$, which is the aforementioned debt, to the $(i+1)$th recursion level.
Observe that this debt does not exceed the out-degree of the examined vertex $w = z^{(i+1)}$ by more than a constant factor, as required.  

There is also a potential call to $\evalf(z)$; recall that $z$ is a shortcut for $z^{(i)}$.
This call should not be confused with the original call to $\evalf(z)$, in which case $z$ is used as a shortcut for $z^{(0)}$.
In what follows we abandon these shortcuts, and write either $z^{(0)}$ or $z^{(i)}$ explicitly, to avoid ambiguity.
Recall that we charged $O(3^{\ell_{z^{(0)}}})$ units of cost out of the costs of the original call to $\evalf(z^{(0)})$ to
the termination cost of epoch $\cE((u,v),l)$ (triggered by the deletion of edge $(u,v)$ from the graph).
Similarly, we charge $O(3^{\ell^*})$ units of cost out of the costs of the new call to $\evalf(z^{(i)})$ to the termination cost of the level-$\ell^*$ epoch $\cE((z^{(i)},w),l)$
(triggered by the deletion of edge $(z^{(i)},w)$ from the matching).
Observe that the only way for the costs of $\evalf(z^{(i)})$ to exceed
$O(3^{\ell^*})$ is due to a call to $\rand(z^{(i)})$. (Indeed, if a call to $\detr(z^{(i)})$ is made, then its cost is bounded by $O(\dout(z^{(i)})) = O(3^{\ell^*})$.)
However, the new call to   $\rand(z^{(i)})$ should not be confused with the original call $\rand(z^{(0)})$, and is analyzed independently of it.
In particular,   the costs of the new call to $\rand(z^{(i)})$ are not viewed as part of the costs of the original call,
and are charged to epochs that are created or terminated as part of the new call to Procedure $\rand$.

Finally, there is   a potential call to $\evalf(w')$. 
We charge $O(3^{\ell_{w'}})$ units of cost out of the costs of this call to the termination cost of the level-$\ell_{w'}$ epoch $\cE((w,w'),l)$
(triggered by the deletion of edge $(w,w')$ from the matching).
Similarly to above, the only way for the costs of $\evalf(w')$ to exceed $O(3^{\ell_{w'}})$ is due to a call to $\rand(w')$. 
However, the costs of the call to $\rand(w')$ are not viewed as part of the costs of the original call to Procedure $\rand$, 
and are charged only to epochs that are created or terminated as part of this new call to Procedure $\rand$.

Summarizing, the costs of the call to $\rand(z^{(i)})$ are re-distributed subject to the requirements of the claim.
First, at most $O(3^{\ell^*})$ units of cost are charged to  the creation and termination costs of the level-$\ell^*$ epoch $\cE((z^{(i)},w),l)$.
Second, at most $O(3^{\ell_{w'}})$ units of cost are charged to the termination cost of the level-$\ell_{w'}$ epoch $\cE((w,w'),l)$.
Finally, the debt left for recursion level $i+1$ (if the procedure proceeds to that level) is bounded by $O(\dout(z^{(i+1)}))$.
The induction step follows.
\QED
\end{proof}
By Claim \ref{conc}, whenever Procedure $\rand$  terminates, no debt whatsoever is left.
Consequently, the entire cost of this procedure (over all recursion levels) can be re-distributed to various epochs
in a manner satisfying the conditions of  Lemma \ref{basiccost}. 
This completes the proof of Lemma \ref{basiccost}.
\QED
\end{proof}
\vspace{1pt}
\noindent
{\bf 4.4~ Natural versus induced epochs.~}
An epoch corresponding to edge $(u,v)$ is terminated either because edge $(u,v)$ is deleted from the graph, and then it is called a \emph{natural epoch},
or because the update algorithm deleted edge $(u,v)$ from the matching, and then it is called an \emph{induced epoch}.
Following \cite{BGS11}, we will charge the computation cost of each induced epoch $\cE$ of level $\ell \ge 0$ to the cost of the unique epoch $\cE'$ whose creation ``triggered'' the termination of
$\cE$.

An induced epoch is terminated only by Procedure $\rand$. Consider the first call to Procedure $\rand(z)$, which we also view as the 0th recursion level.
First, some vertex $w$ of level $\ell_w$ lower than $\ell^*$ is chosen as a random mate for   vertex $z$, and a level-$\ell^*$ epoch $\cE((z,w),l)$ is created.
If $w$ is matched to $w'$, the edge $(w,w')$ is deleted from the matching, and we view the creation of the level-$\ell^*$ epoch $\cE((z,w),l)$ as terminating the
level-$\ell_{w'}$ epoch $\cE((w,w'),l)$.
Next, suppose that
$\dout(w) \ge 3^{\ell^*+1}$. In this case the newly created epoch $\cE((z,w),l)$
is terminated, and immediately afterwards we make the 1st recursive call to $\rand(w)$. 
By Lemma \ref{rand_basic}(1), the call to $\rand(w)$ is guaranteed to create a new epoch $\cE((w,x),l)$ at level $\ell'$ higher than $\ell^*$,
where $x$ is a random outgoing neighbor of $w$, which is of level $\ell_x$ lower than $\ell'$ by Lemma \ref{rand_basic}(2).
We view the creation of the level-$\ell'$ epoch $\cE((w,x),l)$ as terminating the level-$\ell^*$ epoch $\cE((z,w),l)$. Furthermore, if $x$ is matched to $x'$, the edge $(x,x')$ is deleted from the matching,
and we view the creation of the level-$\ell'$ epoch $\cE((w,x),l)$ as terminating the level-$\ell_{x'}$  epoch $\cE((x,x'),l)$ as well.
We have shown that the creation of epoch $\cE((w,x),l)$ at the 1st recursion level terminates (at most) two epochs of lower levels.
Exactly the same reasoning applies to an arbitrary recursion level $i$ by induction.
To summarize: (1) The creation of a new epoch terminates at most two epochs.
(2) The levels of the terminated epochs are strictly lower than that of the created epoch,
i.e., if the level of the created epoch is $\ell$ and the levels of the terminated epochs are $\ell_1$ and $\ell_2$,
then $\ell_1,\ell_2 < \ell$.

We henceforth define the \emph{recursive cost} of an epoch as the sum of its actual cost and the recursive costs of the (at most) two induced epochs terminated by it;
thus the recursive cost of a level-0 epoch is its actual cost.
Denote the highest possible recursive cost of a level-$\ell$ epoch by $\hat C_\ell$, for any $\ell \ge 0$. (Obviously $\hat C_\ell$ is monotone non-decreasing with $\ell$.)
By Lemma \ref{basiccost}, we obtain the recurrence $\hat C_{\ell} \le  \hat C_{\ell_1} +  \hat C_{\ell_2} + O(3^{\ell}) \le  2 \hat C_{\ell-1} + O(3^{\ell})$,
with the base condition
$\hat C_{0} =  O(1)$. This recurrence resolves to $\hat C_{\ell} = O(3^{\ell})$.

\vspace{-2.5pt}
\begin{corollary} \label{important}
For any $\ell \ge 0$, the recursive cost of any level-$\ell$ epoch is bounded by $O(3^{\ell})$.
\end{corollary}
\vspace{-2pt}

By definition, the sum of recursive costs over all natural epochs is equal to the sum of actual costs over all  epochs (both natural and induced)
that have been \emph{terminated} throughout the update sequence.
\vspace{7pt}
\\
\noindent
{\bf 4.5~ Bounding the algorithm's runtime.~}
During any sequence of $t$ updates, the total number of epochs created equals the number of epochs terminated and the number of epochs
that \emph{remain alive} at the end of the $t$ updates.
To bound the computation cost charged to all epochs that remain alive at the end of the update sequence,
one may employ an argument similar to \cite{BGS11}.
However, instead of distinguishing between  epochs that have been terminated and ones that remain alive,
we find it more elegant to get rid of those epochs that remain alive by deleting all edges of the final graph, one after another.
That is, we append  additional edge deletions at the end of the original update sequence, so as to finish with an empty graph.
The order in which these edges are deleted from the graph may be random, but it may also be deterministic,
as long as it is oblivious to the maintained matching; e.g., it can be determined deterministically according to some lexicographic rules that are fixed at the outset of the algorithm.

This tweak guarantees that no epoch remains alive at the end of the update sequence.
As a result, the sum of recursive costs over all natural epochs will bound the sum of actual costs over all  epochs (both natural and induced)
that have been \emph{created} (and also terminated) throughout the update sequence, or in other words, it will bound the total runtime of the algorithm.
Note that the total runtime of the algorithm may only increase as a result of this tweak.
Since we increase the length of the update sequence by at most a factor of 2,
an amortized runtime bound of the algorithm with respect to the new update sequence will imply the same (up to a factor of 2) amortized bound with respect to the original sequence.

\ignore{
consider a level-$\ell$ epoch  corresponding to edge $(u,v)$ that belongs to the matching at the end of the $t$ update steps, and suppose that $u$ initiated the creation of this epoch
at update step $l$.
By Corollary \ref{important}, the recursive cost charged to that epoch is $O(3^\ell)$.
By Corollary \ref{Ob:basic}(1), if $\cO^l_u$ is the set of $u$'s outgoing neighbors at step $l$,
then $|\cO^l_u| \ge 3^\ell$. Note also that $|\cO^l_u| \le t_u$, where $t_u$ is the number of edge updates ever incident on $u$.
Since any vertex may be part of at most one epoch that is alive at the end of the update sequence, the total computation cost charged to all such epochs
is bounded by $\sum_{u \in V} O(t_u) = O(t)$.
}

Define $Y$ to be the r.v.\ for the sum of recursive costs over all natural epochs terminated during the entire sequence.
In light of the above, $Y$ stands for the total runtime of the algorithm.
The proof of the next lemma follows similar lines as in \cite{BGS11}, and is given mainly for completeness.
Nevertheless, since our algorithm is inherently different than the BGS algorithm and as other parts in the analysis are different,
the bounds provided by this lemma shave logarithmic factors from the corresponding bounds of \cite{BGS11}.
\begin{lemma} [Proof in App.\ \ref{app:prob}] \label{expect}
(1) $\Expect(Y) = O(t)$. (2) $Y = O(t + n \log n)$ w.h.p.
\end{lemma}
Shaving the $O(n \log n)$ term from the high probability bound requires additional new ideas; see App.\ \ref{app:improve}.

Finally, the space usage of our algorithm is linear in the dynamic number of edges in the graph.
\begin{theorem}
Starting from an empty graph on $n$ fixed vertices, a maximal matching (and thus 2-MCM and also 2-MCVC) can be maintained over any sequence of $t$ edge insertions and deletions
in $O(t)$ time in expectation and w.h.p., and using $O(n+m)$ space, where $m$ denotes the dynamic number of edges.
\end{theorem}

\ignore{
\section{Discussion and Open Questions}
I suggest to say that it would be interesting to obtain worst-case bounds, and that we believe that our ideas can lead to $2+\eps$-MCM
with roughly the same update time $O(f(\eps))$, but we don't know how to achieve optimality.
It's also reasonable not to have this section, so that it seems that everything is closed
}
\clearpage
\noindent
{\bf Acknowledgements.~}
The author is grateful to Amir Abboud, Surender Baswana, Manoj Gupta, David Peleg, Sandeep Sen, Noam Solomon  and Virginia Vassilevska Williams for helpful discussions.
\ignore{
\section*{Acknowledgements}
The author is grateful to Amir Abboud, David Peleg and Virginia Vassilevska Williams for helpful discussions.
}




\clearpage
\pagenumbering{roman}
\appendix
\centerline{\LARGE\bf Appendix}

\section{Tables and Pseudocode} \label{app:tab}

\begin{table*}[thb]
\begin{center}
\resizebox {\textwidth }{!}{
\footnotesize
\begin{tabular}{|c||c|c|c|}
\hline Reference & Update time   & Space & Bound  \\
\hline
\hline
The na\"{\i}ve algorithm & $O(n)$ & $O(n+m)$ & deterministic \\
\hline
Ivkovi\'{c} and Lloyd (WG'93) \cite{IL93} & $O((n+m)^{\frac{\sqrt{2}}{2}})$ & " & " \\
\hline
Neiman and Solomon (STOC'13) \cite{NS13}  & $O(\sqrt{m})$ & " & "    \\
\hline
\hline
Neiman and Solomon (STOC'13) \cite{NS13}  & $O(\inf_{\beta > 1}(\alpha \beta +\log_{\beta}n))$ & " & "    \\
for $\alpha = O(1)$ & $O(\log n/ \log\log n)$ & " & " \\
\hline He et al.\ (ISAAC'14) \cite{HTZ14} &  $O(\alpha + \sqrt{\alpha \log n})$ & " & "   \\
 for $\alpha = O(1)$   &  $O(\sqrt{\log n})$ & " & "   \\
\hline
\hline Baswana et al. (FOCS'11) \cite{BGS11} &   $O(\log n)$ & $O(n \log n + m)$ & expected    \\
    " &    $O(\log n + (n \log^2 n)/t)$ & " & w.h.p.\    \\
\hline
\hline {\bf  This paper} &   {\boldmath $O(1)$} & {\boldmath $O(n+m)$} & {\bf expected and w.h.p.}  \\
\hline
\end{tabular}
}
\end{center}
\vspace{-0.1in}
\caption[]{ \label{tab1} \footnotesize A   comparison of   previous and our results
for dynamic maximal matchings.
The parameter $t$ (used for describing the result of \cite{BGS11}) designates the total number of updates
and the parameter $\alpha$ (used for describing the results of \cite{NS13,HTZ14}) designates the \emph{arboricity}
bound of the dynamic graph, i.e., it is assumed that the dynamic graph has arboricity at most $\alpha$ at all times.
(The \emph{arboricity} $\alpha(G)$ of a graph $G$ is the minimum number of edge-disjoint forests into which it can be partitioned,
and it is close to the density of its densest subgraph; for any $m$-edge graph, its arboricity ranges between 1 and $\sqrt{m}$.)
}
\end{table*}

\begin{table*}[thb]
\begin{center}
\resizebox {\textwidth }{!}{
\footnotesize
\begin{tabular}{|c||c|c|c|}
\hline Reference & Approximation & Update time   & Bound  \\
\hline
\hline Bhattacharya et al.\ (SODA'15)   \cite{BHI15} & $3+\eps$ & $O(\min\{\sqrt{n}/\eps,m^{1/3}\cdot \eps^{-2}\})$ & deterministic  \\
\hline Bernstein and Stein (SODA'16)   \cite{BS16} & $3/2+\eps$ & $O(m^{1/4} \cdot \eps^{-2.5})$ & " \\
\hline
Neiman and Solomon (STOC'13) \cite{NS13}  & $3/2$ & $O(\sqrt{m})$ & "    \\
\hline Gupta and Peng (FOCS'13) \cite{GP13}  & $1+\eps$ & $O(\sqrt{m} \cdot \eps^{-2})$ &  "   \\
\hline
\hline Gupta and Sharma \cite{GS09} (for trees) & 1 & $O(\log n)$ & " \\
\hline
\hline Bernstein and Stein (SODA'16)  \cite{BS16} & $3/2+\eps$ & $O(\alpha(\alpha + \log n + \eps^{-2}) + \eps^{-6})$ & "  \\
\hline Peleg and Solomon (SODA'16) \cite{PS16} & $3/2+\eps$ & $O(\alpha/\eps)$ & " \\
\hline Bernstein and Stein (ICALP'15)  \cite{BS15} & $1+\eps$ & $O(\alpha(\alpha + \log n) + $ & "  \\
(bipartite graphs of bounded arboricity) & &  $\eps^{-4}(\alpha + \log n) + \eps^{-6})$ & \\
\hline Peleg and Solomon (SODA'16) \cite{PS16} & $1+\eps$ & $O(\alpha \cdot \eps^{-2})$ & " \\
\hline
\hline Neiman and Solomon (STOC'13) \cite{NS13}  & 2 & $O(\inf_{\beta > 1}(\alpha \beta +\log_{\beta}n))$ & "    \\
\hline He et al.\ (ISAAC'14) \cite{KKPS14} & " &  $O(\alpha + \sqrt{\alpha \log n})$ & "  \\
\hline
\hline Onak and Rubinfeld \cite{OR10} & $c$ & $O(\log^2 n)$ & w.h.p. \\
\hline Baswana et al. (FOCS'11) \cite{BGS11} & 2  &   $O(\log n)$ & expected  \\
    " & "  & $O(\log n + (n \log^2 n)/t)$ &  w.h.p. \\
\hline
\hline {\bf  This paper} & {\boldmath $2$} & {\boldmath $O(1)$} & {\bf expected and w.h.p.} \\
\hline
\end{tabular}
}
\end{center}
\vspace{-0.1in}
\caption[]{ \label{tab2} \footnotesize A comparison of previous and our results
for dynamic approximate MCMs. The parameter $t$ designates the total number of updates,
$\alpha$ designates the arboricity bound of the dynamic graph and $c$ is a sufficiently large constant.
In contrast to all other results, the matching maintained in \cite{GS09} is not represented \emph{explicitly}: In order to determine if an edge belongs to the matching, one should run an $O(\log n)$-time query. This result appears in the table for completeness.
}
\end{table*}

\begin{table*}[thb]
\begin{center}
\resizebox {\textwidth }{!}{
\footnotesize
\begin{tabular}{|c||c|c|c|}
\hline Reference & Approximation & Update time   & Bound  \\
\hline \hline
Neiman and Solomon (STOC'13) \cite{NS13}  & $2$ & $O(\sqrt{m})$ & deterministic   \\
\hline
\hline Neiman and Solomon (STOC'13) \cite{NS13}  & 2 & $O(\inf_{\beta > 1}(\alpha \beta +\log_{\beta}n))$ & "  \\
\hline He et al.\ (ISAAC'14) \cite{KKPS14} & 2 &  $O(\alpha + \sqrt{\alpha \log n})$ & "  \\
  \hline
\hline Bhattacharya et al.\ (SODA'15)   \cite{BHI15} & $2+\eps$ & $O(\log n \cdot \eps^{-2}\})$ & " \\
\hline
\hline Peleg and Solomon (SODA'16) \cite{PS16} &   $2+\eps$ & $O(\alpha/\eps)$ & "  \\
  \hline
  \hline Onak and Rubinfeld \cite{OR10} & $c$ & $O(\log^2 n)$ & w.h.p. \\
  \hline Baswana et al. (FOCS'11) \cite{BGS11} & 2  & expected $O(\log n)$ & expected  \\
    " & " & $O(\log n + (n \log^2 n)/t)$ & w.h.p.  \\
    \hline
\hline {\bf  This paper} & {\boldmath $2$} & {\boldmath $O(1)$} & {\bf expected and w.h.p.} \\
\hline
\end{tabular}
}
\end{center}
\vspace{-0.1in}
\caption[]{ \label{tab3} \footnotesize A  comparison of previous and our results for dynamic approximate MCVCs.
The parameter $t$ designates the total number of updates,
$\alpha$ designates the arboricity bound of the dynamic graph and $c$ is a sufficiently large constant.
}
\end{table*}


\begin{figure}[!ht]
\fbox{
\begin{minipage}[t]{165mm}
$\ins(u,v)$:
\begin{enumerate}\itemsep1.5pt \parskip2pt \parsep2pt
\item $N_v \leftarrow N_v \cup \{u\}$;
\item $N_u \leftarrow N_u \cup \{v\}$;
\item If $\ell_u \ge \ell_v$:
 /* orient the edge from $u$ to $v$ */
   \begin{enumerate}\itemsep1.5pt \parskip2pt \parsep2pt
   \item $\cO_u \leftarrow \cO_u \cup \{v\}$;
   \item $\cI_v[\ell_u] \leftarrow \cI_v[\ell_u] \cup \{u\}$;
   \end{enumerate}
\item Else:
 /* orient the edge from $v$ to $u$ */
   \begin{enumerate}\itemsep1.5pt \parskip2pt \parsep2pt
   \item $\cO_v \leftarrow \cO_v \cup \{u\}$;
   \item $\cI_u[\ell_v] \leftarrow \cI_u[\ell_v] \cup \{v\}$;
   \end{enumerate}
\item If $\ell_u = \ell_v = -1$:  /* if $u$ and $v$ are free, match them */
   \begin{enumerate}\itemsep1.5pt \parskip2pt \parsep2pt
   \item  $M \leftarrow M \cup \{(u,v)\}$;
   \item $\setlvl(u,0)$;
   \item $\setlvl(v,0)$;
   \end{enumerate}
\end{enumerate}
\end{minipage}}
\caption{Handling edge insertion $(u,v)$.
}
\label{fig:handleIns}
\end{figure}

\begin{figure}[!ht]
\fbox{
\begin{minipage}[t]{165mm}
$\del(u,v)$:
\begin{enumerate}\itemsep2pt \parskip2pt \parsep2pt
\item $N_v \leftarrow N_v \setminus \{u\}$;
\item $N_u \leftarrow N_u \setminus \{v\}$;
\item If $v \in \cO_u$:
 /* if edge $(u,v)$ is oriented from $u$ to $v$ */
   \begin{enumerate}\itemsep2pt \parskip2pt \parsep2pt
   \item $\cO_u \leftarrow \cO_u \setminus \{v\}$;
   \item $\cI_v[\ell_u] \leftarrow \cI_v[\ell_u] \setminus \{u\}$;
   \end{enumerate}
\item Else:
   \begin{enumerate}\itemsep2pt \parskip2pt \parsep2pt
   \item $\cO_v \leftarrow \cO_v \setminus \{u\}$;
   \item $\cI_u[\ell_v] \leftarrow \cI_u[\ell_v] \setminus \{v\}$;
   \end{enumerate}
\item If $(u,v) \in M$:
 /* delete edge $(u,v)$ from the matching */
   \begin{enumerate}\itemsep2pt \parskip2pt \parsep2pt
   \item  $M \leftarrow M \setminus \{(u,v)\}$; /* $u$ and $v$ become temporarily free, their levels exceed -1 */
   \item $\evalf(u)$;
   \item $\evalf(v)$;
   \end{enumerate}
\end{enumerate}
\end{minipage}}
\caption{Handling edge deletion $(u,v)$.
}
\label{fig:handleDel}
\end{figure}

\begin{figure}[!ht]
\fbox{
\begin{minipage}[t]{165mm}
$\setlvl(v,\ell)$:
\begin{enumerate}\itemsep2pt \parskip2pt \parsep2pt
    \item For all $w \in \cO_v$:  /* update $\cI_w$ regarding $v$'s new level  */
       \begin{enumerate}\itemsep2pt \parskip2pt \parsep2pt
        \item $\cI_w[\ell_v] \leftarrow \cI_w[\ell_v] \setminus \{v\}$;
        \item $\cI_w[\ell] \leftarrow \cI_w[\ell] \cup \{v\}$;
        \end{enumerate}

\item If $\ell < \ell_v$: /* in this case the level of $v$ is decreased by at least one */
    \begin{enumerate}\itemsep2pt \parskip2pt \parsep2pt
    \item For all $w \in \cO_v$ such that $\ell+1 \le \ell_w \le \ell_v$:
     /* flip $v$'s outgoing edge $(v,w)$ */
       \begin{enumerate}\itemsep2pt \parskip2pt \parsep2pt
        \item $\cO_v \leftarrow \cO_v \setminus \{w\}$;
        \item $\cI_v[\ell_w] \leftarrow \cI_v[\ell_w] \cup \{w\}$;
        \item $\cI_w[\ell] \leftarrow \cI_w[\ell] \setminus \{v\}$;
        \item $\cO_w \leftarrow \cO_w \cup \{v\}$;
        \end{enumerate}
   \end{enumerate}

\item If $\ell > \ell_v$: /* in this case the level of $v$ is increased by at least one */
 \begin{enumerate}\itemsep2pt \parskip2pt \parsep2pt
    \item For all $i = \ell_v,\ldots,\ell-1$ and all $w \in \cI_v[i]$:
   /* flip $v$'s incoming edge $(v,w)$ */
       \begin{enumerate}\itemsep2pt \parskip2pt \parsep2pt
       \item $\cI_v[i] \leftarrow \cI_v[i] \setminus \{w\}$;
       \item $\cO_v \leftarrow \cO_v \cup \{w\}$;
       \item $\cO_w \leftarrow \cO_w \setminus \{v\}$;
       \item $\cI_w[\ell] \leftarrow \cI_w[\ell] \cup \{v\}$;
       \end{enumerate}
\end{enumerate}
\item  $\ell_v \leftarrow \ell$;
\end{enumerate}
\end{minipage}}
\caption{Setting the old level $\ell_v$ of $v$ to   $\ell$.}
\label{fig:setlvl}
\end{figure}

\begin{figure}[!ht]
\fbox{
\begin{minipage}[t]{165mm}
$\evalf(v)$:
\begin{enumerate}\itemsep2pt \parskip2pt \parsep2pt
\item If $\dout(v)  < 3^{\ell_v+1}$: $\detr(v)$;
\item Else $\rand(v)$;
\end{enumerate}
\end{minipage}}
\caption{Handling a vertex that becomes temporarily free.
}
\label{fig:evalfree}
\end{figure}

\begin{figure}[!ht]
\fbox{
\begin{minipage}[t]{165mm}
$\detr(v)$:
\begin{enumerate}\itemsep2pt \parskip2pt \parsep2pt
\item For all $w \in \cO_v$:
       \begin{enumerate}\itemsep2pt \parskip2pt \parsep2pt
        \item If $\ell_w = -1$: /* if $w$ is free, match $v$ with $w$ */
        \begin{enumerate}\itemsep2pt \parskip2pt \parsep2pt
        \item  $M \leftarrow M \cup \{(v,w)\}$;
        \item $\setlvl(v,0)$;
        \item $\setlvl(w,0)$;
        \item terminate;
        \end{enumerate}
   \end{enumerate}
 \item \setlvl(v,-1); /* all  outgoing neighbors of $v$ are matched, hence $v$ becomes free */
\end{enumerate}
\end{minipage}}
\caption{Matching $v$ with a free neighbor  (if exists) deterministically. 
It is assumed that $\dout(v)  < 3^{\ell_v+1}$.
}
\label{fig:detr}
\end{figure}

\begin{figure}[!ht]
\fbox{
\begin{minipage}[t]{165mm}
$\rand(v)$:
\begin{enumerate}\itemsep2pt \parskip2pt \parsep2pt
\item $\ell^* \leftarrow \ell_v$;
\item while $\phi_v(\ell^*+1) \ge 3^{\ell^*+1}$: $\ell^* \leftarrow \ell^* + 1$;
\\ /* $\ell^*$ is the minimum level after $\ell_v$ with $\phi_v(\ell^*+1) < 3^{\ell^*+1}$ */
\item $\setlvl(v,\ell^*)$; /* after this call $\ell_v = \ell^*$ and $3^{\ell^*} \le \dout(v) = \phi_v(\ell^*) < 3^{\ell^* + 1}$ */

\item Pick an outgoing neighbor $w$ of $v$ uniformly at random;
\\ /* $w$ is chosen with probability at most $1/3^{\ell^*}$ and $\ell_w \le \ell^* -1$ */
\item $w' \leftarrow \mate(w)$;
\item If $w' \ne \bot$: $M \leftarrow M \setminus \{(w,w')\}$;
\item $\setlvl(w,\ell^*)$; /* in order to match $v$ to $w$, they need to be at the same level */
\item $M \leftarrow M \cup \{(v,w)\}$;
\item If $\dout(w) \ge  3^{\ell^*+1}$:
    \begin{enumerate}\itemsep2pt \parskip2pt \parsep2pt
    \item $M \leftarrow M \setminus \{(v,w)\}$; /* after this command is executed, $\mate(v) = \mate(w) = \bot$ */
    \item $\rand(w)$; /* before this call, $\dout(w) \ge  3^{\ell^*+1} = 3^{\ell_w+1}$ */
    \item If $\mate(v) = \bot$: $\evalf(v)$; /* if $v$ is temporarily free, handle it */
    \end{enumerate}
\item If $w' \ne \bot$ and $\mate(w') = \bot$: $\evalf(w')$; /* if $w'$ is temporarily free, handle it */
\end{enumerate}
\end{minipage}}
\caption{Matching $v$ at level $\ell^*$ higher than $\ell_v$, with a random neighbor $w$ of level lower than $\ell^*$.
If this   requires too much time, the procedure calls itself recursively to match $w$ at yet a higher level, in which case $v$ becomes free and handled via Procedure $\evalf(v)$.
It is assumed that $\dout(v)  \ge 3^{\ell_v+1}$.
}
\label{fig:rand}
\end{figure}


\ignore{
\section{Some More Details on the Data Structures} \label{app:data}
While the number of pointers stored in the  dynamic hash table $\cI_v$ is bounded by $\log_3 (n-1)+2$, it may be much smaller than that.
Indeed, there is no pointer in $\cI_v$ corresponding to level smaller than $\ell_v$,
implying that the number of pointers stored in $\cI_v$ is at most $\log_3 (n-1) + 1 -\ell_v$.
Note also that no incoming neighbor of $v$ is of level -1 (by Invariant \ref{exception}), thus there is no pointer in $\cI_v$ corresponding to level -1.
Recall that pointers to empty lists are not stored in the hash table, thus the total space over all hash tables is at most linear in the dynamic number of edges
in the graph. If we were to store the pointers in a static array, then the total space over these arrays would be $\Omega(n \log n)$.
In the case that the dynamic graph is usually dense (i.e., having $\Omega(n \log n)$ edges), it is advantageous to use arrays,
as it is easier to implement all the basic operations (delete, insert, search), and the time bounds become deterministic.

%
%
%
%

Note that no information whatsoever on the levels of $v$'s outgoing neighbors is provided by the data structure $\cO_v$.
In particular, to determine if $v$ has an outgoing neighbor at a certain level (most importantly at level -1, i.e., a free neighbor),
we need to scan the entire list $\cO_v$. On the other hand, $v$ has an incoming neighbor at a certain level $\ell$ iff
the corresponding list $\cI_v[\ell]$ is non-empty.
It will not be in $v$'s \emph{responsibility} to maintain the
data structure $\cI_v$, but rather within the responsibility of $v$'s incoming neighbors.

As mentioned, we keep mutual pointers between the elements in the various data structures: For any vertex $v$ and any outgoing neighbor $u$ of $v$,
we have mutual pointers between all elements $u \in \cO_v, v \in \cI_u[\ell_v],v \in N_u,u \in N_v$.
For example, when  an edge $(u,v)$ oriented as ${\orient v u}$ is deleted from the graph,
we get a pointer  to either $v \in N_u$ or $u \in N_v$, and through this pointer we delete all elements $u \in \cO_v, v \in \cI_u[\ell_v],v \in N_u,u \in N_v$.
As another example, when  the orientation of edge $(u,v)$ is flipped from ${\orient u v}$ to ${\orient v u}$, then assuming we have
a pointer to   $v \in \cO_u$ (which is   the case in our algorithms), we can reach element $u \in \cI_v[\ell_u]$ through the pointer, delete them both from the respective lists,
and then create elements $u \in \cO_v$ and $v \in \cI_u[\ell_v]$ with mutual pointers.
We also keep mutual pointers between a matched edge  and its endpoints.
(We do not provide a complete description of the straightforward yet tedious maintenance of these pointers for the sake of brevity.)
}


\section{Proof  of Lemma \ref{expect}} \label{app:prob}
{
While both endpoints of a matched edge are of the same level, this may not be the case for an unmatched edge.
We say that an edge $e$ is \emph{deleted at level} $\ell$ (from the graph) if \emph{at least one} of its endpoints is at level $\ell$
at the time of the deletion.
(Thus each edge is deleted at either one or two levels.
While the same edge can be deleted and inserted multiple times in an update sequence, we view different occurrences of the same edge as different objects.)
Let $S_\ell$ denote the sequence of edge deletions at level $\ell$,   write $|S_\ell| = t_\ell$, and denote by $t_{del}$ the total number of deletions;
Note that $t_\ell$ is a random variable (r.v.).
Then $\sum_{\ell \ge 0} t_\ell \le 2t_{del} \le t$.
Let $X_\ell = X_\ell(S_\ell)$ be the r.v.  for the number of natural epochs terminated at level $\ell$ for the update sequence $S_\ell$,
and let $Y_\ell$ be the r.v.\ for the sum of recursive costs over these epochs.
By Corollary \ref{important}, $Y_{\ell} = O(3^\ell) \cdot X_\ell$.
Recall that $Y$ is the r.v.\ for the sum of recursive costs over all natural epochs terminated during the entire sequence.
Thus we have $Y = \sum_{\ell \ge 0} Y_\ell$.  
}

Fix an arbitrary level $\ell, 0 \le \ell \le \log_3 (n-1)$.
Consider any natural level-$\ell$ epoch initiated by some vertex $u$ at some update step $l$,
and let $\cO^{l}_{u}$ denote the set of $u$'s outgoing neighbors at that time.
By Corollary \ref{Ob:basic}, $|\cO^{l}_{u}| \ge 3^\ell$,
and the mate $v$ of $u$ is chosen uniformly at random among all vertices of  $\cO^l_u$.
The epoch is terminated when the matched edge $(u,v)$ gets deleted \emph{from the graph};
we define the \emph{(regular) duration of the epoch} as the number of outgoing edges of $u$ at time $l$
that get deleted  from the graph between time step $l$ and the epoch's termination.
(All these edge deletions occur at level $\ell$ by definition.)
\begin{observation} \label{obs:duration}
If there are $q$ natural level-$\ell$ epochs with durations at least $\delta$,
then $q \le 2t_\ell / \delta$.
\end{observation}
\begin{proof}
Consider the edge deletions that define the durations of these $q$ epochs.
Any such edge deletion $(u,v)$ is associated with at most two epochs, one initiated by $u$ and possibly another one initiated by the other endpoint $v$.
Hence the total number of such deletions is bounded by $2t_\ell$, and we are done.
\QED
\end{proof}
\vspace{2pt}
\noindent
\\
{\bf Proof of   Lemma \ref{expect}(1):}~
We say that a level-$\ell$ epoch is \emph{short} if its duration is at most $(1/2) 3^{\ell}$;
otherwise it is \emph{long}.
By definition, when a short epoch is terminated, at least half of the   edges among which the matched edge was randomly chosen   are still present in the graph.
Let $X^{short}_\ell$ and $X^{long}_\ell$ be the random variables for the number of short and long epochs terminated at level $\ell$, respectively.
By definition, we have $X_\ell = X^{short}_\ell + X^{long}_\ell$. By Observation \ref{obs:duration}, $X^{long}_\ell \le 2t_\ell / (3^{\ell}/2) = 4t_\ell / 3^\ell$.

Let $e = (u,v)$ be an edge deleted (from the graph) at level $\ell$ during update step $l$, and let $Z_e$ be the indicator random variable
that takes value 1 if the deletion of edge $e$ causes termination of a \emph{short} epoch at level $\ell$, and 0 otherwise.
Observe that $X^{short}_\ell = \sum_{e \in S_\ell} Z_e$.
\ignore{
defined as follows.
\begin{equation*}
  Z_e=\begin{cases}
    1, & \text{if the deletion of edge $e$ causes termination of an epoch at level $i$}.\\
    0, & \text{otherwise}.
  \end{cases}
\end{equation*}
}

Suppose that the deletion of edge $e$ causes termination of a short epoch at level $\ell$,
and assume w.l.o.g.\ that $u$ was the initiator of this epoch. Let $l'$ be the update step at which the corresponding epoch $\cE(e,l')$
was created, and let $\cO^{l'}_{u}$ be the set of $u$'s outgoing neighbors at that time.
By Corollary \ref{Ob:basic}, $|\cO^{l'}_{u}| \ge 3^\ell$.
Moreover, $u$ selects $v$ as its mate at update step $l'$ uniformly at random among all vertices of $\cO^{l'}_u$.
By Invariant \ref{exception}(c), the level of edge $e$ remains $\ell$ throughout the epoch's existence,
so if any of $u$'s outgoing edges at update step $l'$ is deleted during this time interval,  it is deleted at level $\ell$ by definition.

We need to bound the probability  that the deletion of edge $e$ at update step $l$ causes termination of a short epoch at level $\ell$,
given that this epoch has not been terminated yet.
Since this epoch is short, it suffices to bound the probability that edge $e$ was chosen at step $l'$ among the at least $|\cO^{l'}_{u}|/2$
edges that are still present in the graph. We know that each of these edges has the same probability of being chosen, hence
this probability is bounded by $2/|\cO^{l'}_{u}|$,
and we have $\Prob(Z_e = 1) \le 2 / |\cO^{l'}_u| \le 2/3^\ell$.
Note also that $\sum_{\ell \ge 0} t_\ell \le t$, which implies that $\sum_{\ell \ge 0} \Expect(t_\ell) = \Expect(\sum_{\ell \ge 0} t_\ell) \le \Expect(t) = t$.
It follows that
\begin{eqnarray*}
\Expect(Y) &=& \sum_{\ell \ge 0} \Expect(Y_\ell) ~=~
\sum_{\ell \ge 0} O(3^{\ell}) \cdot \Expect(X_\ell) ~=~
\sum_{\ell \ge 0} O(3^{\ell}) \cdot (\Expect(X^{short}_\ell) + \Expect(X^{long}_\ell)) \\ &\le&
\sum_{\ell \ge 0} O(3^{\ell}) \cdot \left(\left(\sum_{e \in S_\ell} \Expect[Z_e] \right) + \Expect(X^{long}_\ell)\right)
~=~ \sum_{\ell \ge 0} O(3^{\ell}) \cdot \left(\left(\sum_{e \in S_\ell}  \Prob[Z_e = 1] \right) + (4/3^\ell) \Expect(t_\ell) \right)
\\  &\le& \sum_{\ell \ge 0} O(3^{\ell}) \cdot \left(\left(\sum_{e \in S_\ell} (2/3^\ell)\right) + (4/3^\ell) \Expect(t_\ell) \right)
~\le~ O(1) \cdot  \sum_{\ell \ge 0} \left(2t_\ell + 4\Expect(t_\ell)\right) ~=~  O(t). \inQED
\end{eqnarray*}
\ignore{


By linearity of expectation,
$\Expect(X^{short}_\ell) ~=~ \sum_{e \in S_\ell} \Expect[Z_e] ~=~ \sum_{e \in S_\ell} \Prob[Z_e = 1] ~\le~ \sum_{e \in S_\ell} (2/3^\ell)$.
It follows that $$Y_\ell ~=~ O(3^{\ell}) \cdot X_\ell ~=~ O(3^{\ell}) \cdot (X^{short}_\ell + X^{long}_\ell) ~\le~  O(3^{\ell}) \cdot (3t_\ell / 3^{\ell})
~=~ O(t_\ell),$$ from which we conclude
}
\vspace{8pt}
\noindent
\\
{\bf Proof of Lemma \ref{expect}(2):}~
\ignore{
By Corollary \ref{Ob:basic}, $|\cO^{l}_{u}| \ge 3^\ell$,
and the mate $v$ of $u$ is chosen uniformly at random among all vertices of  $\cO^l_u$.
The epoch is terminated when the matched edge $(u,v)$ gets deleted (from the graph); we define the \emph{duration of the epoch} as the number of outgoing edges of $u$ at time $l$
that get deleted  between time step $l$ and the epoch's termination.
(All these edge deletions occur at level $\ell$ by definition.)
}
Fix any level $\ell, 0 \le \ell \le \log_3 (n-1)$,
consider a   level-$\ell$ epoch initiated by some vertex $u$ at some update step $l$,
and let $\cO^{l}_{u}$ denote the set of $u$'s outgoing neighbors at that time.
We define the \emph{uninterrupted duration} of the epoch as the number of outgoing edges of $u$ at time $l$
that get deleted from the graph between time step $l$ and the time that the random matched edge $(u,v)$ is deleted from the graph.
(Since we appended edge deletions at the end of the original update sequence to guarantee that the final graph is empty,
all the outgoing edges of $u$ at time $l$, including edge $(u,v)$, will get deleted throughout the update sequence.)
If the epoch is natural, then the uninterrupted duration of an epoch is equal to its (regular) duration. However, for an induced epoch,
its uninterrupted duration may be significantly larger than its duration.
(Note also that the outgoing edges of $u$ at time $l$ that get deleted from the graph after the epoch's termination are not necessarily deleted at level $\ell$.)

We argue that the epoch's uninterrupted duration
is a r.v.\ uniformly distributed in the range $[1,|\cO^l_u|]$.
\begin{claim} \label{eta}
For any $1 \le k \le |\cO^l_u|$, the probability that the uninterrupted duration of the epoch is
precisely $k$ equals $1 / |\cO^l_u| \le 1/3^\ell$.
This bound remains valid even if it is given that the level of the epoch is $\ell$.
 \end{claim}
\begin{proof}
Denote the outgoing edges of $u$ at time $l$ by $e_1,e_2, \ldots, e_\rho$, where $\rho = |\cO^l_u|$,
let $d_1,d_2,\ldots,d_\rho$ denote the times at which these edges are deleted from the graph, respectively, and assume w.l.o.g.\ that $d_1 < d_2 < \ldots d_\rho$.
Denote the edge associated with the epoch by $e_i$.
By time $d_i$, all edges $e_1,\ldots,e_{i}$ have been deleted from the graph,
but all edges $e_{i+1},\ldots,e_\rho$   remain there. Consequently, for the uninterrupted duration of the epoch to equal $k$, it must hold that $e_i = e_k$.
Thus the probability that the epoch's uninterrupted duration equals $k$ is given by the probability that its associated edge is $e_k$.
Since this edge is chosen uniformly at random among $\{e_1,\ldots,e_\rho\}$, the probability of choosing $e_k$ equals $1 / \rho = 1/ |\cO^l_u|$.
Moreover, this bound remains valid even if it is given that the epoch's level is $\ell$.
\QED
\end{proof}

Note that the uninterrupted durations of distinct level-$\ell$ epochs are not necessarily independent:
\begin{itemize}
\item
First, the number $|\cO^l_u|$ of outgoing neighbors from which a mate $v$ for $u$ is chosen
may   well depend on previous coin flips of the algorithm. However, by Corollary \ref{Ob:basic},
it must be that $|\cO^l_u| \ge 3^\ell$.
\item Second, although the mate of $u$ is chosen uniformly at random among all vertices in $\cO^l_u$,
some of the optional choices may preclude future events from happening. Hence, future events may depend on this random choice.
Moreover, this choice may well effect the (regular) durations of epochs that were created prior to it.
However, this choice does not effect the uninterrupted durations of such epochs.  
Also, as $|\cO^l_u| \ge 3^\ell$, the probability that a specific vertex $v$ in $\cO^l_u$ is chosen as $u$'s random mate
is bounded by $1/3^\ell$, independently of epochs that were created prior to the current epoch.
\end{itemize}
It follows that
 the probability that the uninterrupted duration of a level-$\ell$ epoch
equals $k$ is bounded by $1/3^\ell$, for any $k$,
even if it is given that the level of the epoch is $\ell$, 
and independently of
the uninterrupted durations of  epochs that were created prior to the current epoch.
We derive the following corollary.
\begin{corollary} \label{corborder}
For any $1 \le k \le 3^\ell$, the probability that the epoch's uninterrupted duration is at most $k$ is bounded by $k/3^\ell$,
even if it is given that the level of the epoch is $\ell$, and independently of the uninterrupted durations of epochs that were created prior to the current epoch's creation.
\end{corollary}

Let $T_\ell$ be the r.v.\ for the total number of epochs (both induced and natural) terminated at level $\ell$, and assume that $T_\ell \le 2X_\ell$;
in Section \ref{justify} we demonstrate that this assumption does not lose generality.
(Since we made sure that the final graph is empty, any created epoch will get terminated throughout the update sequence.
Consequently, $T_\ell$ designates the total number of epochs created at level $\ell$ and $\sum_{\ell \ge 0} T_\ell$ designates the total number of epochs created
over all levels throughout the entire update sequence.)

With a slight abuse of notation from the proof of the first assertion of this lemma, we say that a level-$\ell$ epoch is \emph{$\mu$-short} if its \emph{uninterrupted} duration is at most $\mu \cdot 3^{\ell}$, for some parameter $0 \le \mu \le 1$.

Write $\eta = 1/16e$, let $T'_\ell$ be the r.v.\ for the number of level-$\ell$ epochs that are $\eta$-short,
and let $T''_\ell = T_\ell - T'_\ell$ be the r.v.\ for the number of remaining level-$\ell$ epochs, i.e., those that are not $\eta$-short.
Let $A_\ell$ be the event that both $T_\ell > 4\log n$ and $T'_\ell \ge T_\ell / 4$  hold, or equivalently, $4T'_\ell \ge T_\ell > 4 \log n$.
\begin{claim} \label{first1}
$\Prob(A_\ell) \le 8/ (3n^4)$.
\end{claim}
\begin{proof}
Fix two parameters $q$ and $j$, with $j \ge q/4$,
and consider any $q$ level-$\ell$ epochs $E_1,\ldots,E_q$, ordered by their creation time, so that $E_i$ was created before $E_{i+1}$, for each $i = 1,2,\ldots,q-1$.

We argue that the probability that precisely $j$ particular epochs among these $q$ are $\eta$-short is bounded from above by $\eta^j$.
Indeed, by Corollary \ref{corborder}, the probability of an epoch to be $\eta$-short is at most $\eta$,
even if it is given that the level of the epoch is $\ell$, and independently of
the uninterrupted durations of epochs that were creator prior to the current epoch's creation.
Thus, if we denote by $B^{(i)}$  the event that the $i$th epoch among these $j$ is $\eta$-short, for $1 \le i \le j$,
then we have  $\Prob(B^{(i)} ~\vert~ B^{(1)} \cap B^{(2)} \cap \ldots B^{(i-1)}) \le \eta$. (Moreover,
this upper bound of $\eta$ on the probability continues to hold even if it is given that the $i$th epoch among these $q$, as well as any previously created epoch, is at level $\ell$.)
Consequently,
$$\Prob(B^{(1)} \cap B^{(2)} \cap \ldots \cap B^{(j)}) ~=~ \Prob(B^{(1)}) \cdot \Prob(B^{(2)} ~\vert~ B^{(1)}) \cdot \ldots \cdot \Prob(B^{(j)} ~\vert~ B^{(1)} \cap B^{(2)}
\cap \ldots B^{(j-1)}) ~\le~ \eta^j.$$

Next, we argue that $\Prob[T_\ell = q \cap T'_\ell = j] \le {q \choose j} \eta^j$.
Instead of (i) going over all possibilities of choosing $q$ level-$\ell$ epochs among  all-level epochs,
(ii) bounding the probability that each such possibility is chosen and precisely $j$ epochs out of the chosen $q$ are $\eta$-short,
and (iii) taking the sum of all these probabilities, we handle all such possibilities \emph{together}. That is, we restrict our attention to a smaller sample space
that consists of just the $q$ level-$\ell$ epochs, without actually choosing or fixing them among all epochs,
and bound the probability that precisely $j$ of them are $\eta$-short.
Specifically, let $E_1,\ldots,E_q$ denote the $q$ level-$\ell$ epochs,  
and note that there are ${q \choose j}$ possibilities to choose $j$ $\eta$-short epochs among $E_1,\ldots,E_q$.
As we have shown, each such possibility occurs with probability at most $\eta^j$, and the assertion follows.


Noting that ${q \choose j} \le (eq/j)^j \le (4e)^j$ and recalling that $\eta = 1/16e$, we have ${q \choose j} \eta^j \le (1/4)^j$.
Hence
\begin{eqnarray*}
\Prob(A_\ell) &=& \Prob[T_\ell  >  4\log n \cap T'_\ell \ge T_\ell / 4]
~=~ \sum_{q > 4 \log n} \sum_{j = q/4}^q \Prob[T_\ell = q \cap T'_\ell = j]
\\ &\le& \sum_{q > 4 \log n} \sum_{j = q/4}^q {q \choose j} \eta^j
~\le~ \sum_{q > 4 \log n} \sum_{j = q/4}^q (1/4)^j
~\le~
\sum_{q > 4 \log n} 4/3((1/2)^q)
\\ &\le& 8/3 ((1/2)^{4\log n}) ~\le~ 8/ (3n^4). \inQED
\end{eqnarray*}
\end{proof}

Let $c$ be a sufficiently large constant.
\begin{claim} \label{negate}
If $\neg A_{\ell}$, then $Y_\ell < c(t_\ell + 3^\ell \cdot \log n)$.
\end{claim}
\begin{proof}
If $\neg A_\ell$, then either $T_\ell \le 4\log n$ or $T'_\ell < T_\ell / 4$ must hold.

In the former case $X_\ell \le T_\ell \le 4\log n$, and we have $$Y_\ell ~=~ O(3^\ell) \cdot X_\ell ~\le~ O(3^\ell) \cdot  4\log n ~<~ c 3^\ell \cdot   \log n ~\le~ c(t_\ell + 3^\ell \cdot \log n).$$
Next, suppose that $T'_\ell < T_\ell / 4$.
In this case $T''_\ell > 3T_\ell/4$, i.e.,  more than three quarters of the $T_\ell$  epochs that are terminated at level $\ell$ are   not $\eta$-short.
Recall also that we assume that $T_\ell \le 2X_\ell$ (see Section \ref{justify}), i.e., at least half of the $T_\ell$ epochs are natural.
It follows that at least a quarter of the $T_\ell$ epochs that terminated at level $\ell$ are both natural and not $\eta$-short.
Denoting  by $X''_\ell$ the r.v.\ for the number of epochs that are both natural and not $\eta$-short,
we thus have $X''_\ell \ge T_\ell/4 \ge X_\ell / 4$.
Since these $X''_\ell$ epochs are natural, the duration of each of them is equal to its uninterrupted duration, and thus it exceeds $\eta 3^\ell$.
Therefore, Observation \ref{obs:duration} yields $X''_{\ell} < 2t_\ell / (\eta 3^\ell)$.
We conclude that
$$Y_\ell ~=~ O(3^\ell) \cdot X_\ell ~\le~ O(3^\ell) \cdot 4X''_\ell ~\le~ O(3^\ell) \cdot 8t_\ell / (\eta 3^\ell) ~<~ ct_\ell ~<~ c(t_\ell + 3^\ell \cdot \log n). \inQED$$
\end{proof}

Let $A$ be the event that $Y > c(t + (3/2) n\log n)$.
Claim \ref{negate} yields the following corollary.
\begin{corollary} \label{second2}
If $A$, then $A_0 \cup A_1 \cup \ldots A_{\log_3(n-1)}$.
\end{corollary}
\begin{proof}
We assume that $\neg A_0 \cap \neg A_1 \cap \ldots \neg A_{\log_3(n-1)}$ holds, and show that $A$ cannot hold.
Indeed, by Claim \ref{negate}, we have that $Y_\ell < c(t_\ell + 3^\ell \cdot \log n)$ for each $\ell \ge 0$. It follows that
\begin{equation} \label{sumi}
Y ~=~ \sum_{\ell \ge 0} Y_\ell ~<~ \sum_{\ell \ge 0} c(t_\ell + 3^\ell \cdot \log n)  ~\le~ c(t + (3/2) n\log n). \inQED
\end{equation}
\end{proof}

Claim \ref{first1} and Corollary \ref{second2} imply that
\begin{equation} \label{finall}
\Prob(A) ~\le~ \Prob(A_0 \cup A_1 \cup \ldots A_{\log_3(n-1)}) ~\le~ \sum_{\ell \ge 0} \Prob(A_\ell) ~\le~ (\log_3(n-1) + 1) (8/ (3n^4)) ~=~ O(\log n / n^4).
\end{equation}
It follows that $Y$ is upper bounded by $O(t + n\log n)$ with high probability, as required.
\QED
\ignore{
\begin{eqnarray} \label{claim:basic}
\Prob[X_\ell = q] &\le& \sum_{j = q/2}^q {q \choose j} \left(\frac{2t_\ell}{q (3^\ell)}\right)^j ~\le~
\sum_{j = q/2}^q (2e)^j \left(\frac{2t_\ell}{q (3^\ell)}\right)^j
~=~ \sum_{j = q/2}^q \left(\frac{4et_\ell}{q (3^\ell)}\right)^j.
\end{eqnarray}


Set $q_0 = 4(4et_\ell/3^\ell + \log n)$.
For any $q > q_0$, we have $\Prob[X_\ell = q] \le 4/3((1/2)^q)$ by (\ref{claim:basic}).
Therefore $$\Prob[X_\ell > q_0] ~=~ \sum_{q > q_0} \Prob[X_\ell = q] ~\le~ \sum_{q > q_0} 4/3((1/2)^q) ~\le~ 8/3 (1/2)^{q_0} ~<~ 8/3 (1/2)^{4\log n} ~=~  8/ (3n^4).$$
Thus $X_\ell$ is bounded by $O(t_\ell / 3^\ell + \log n)$ with high probability, as required.}

\subsection{Justifying the assumption} \label{justify}
Consider any level $\ell$ where our assumption does not hold, i.e.,  $T_\ell > 2X_\ell$,
whence the number of induced epochs terminated at level $\ell$ exceeds the number of natural epochs terminated at that level.
Next, we show that the computation costs incurred by level $\ell$ can be charged to the computation costs at higher levels,
allowing us to disregard any level where the assumption does not hold in the runtime analysis.

As $T_\ell > 2X_\ell$,
we may define a one-to-one mapping from the natural to the induced epochs, mapping each natural level-$\ell$ epoch to a unique induced epoch at that level.
Then we can (temporarily) charge the costs of any level-$\ell$ natural epoch to the induced epoch to which it is mapped.
In Section 4.3, we defined the recursive cost of  an epoch as the sum of its actual cost and the recursive costs of the at most two induced
epochs (at lower levels) terminated by it. Let us re-define the recursive cost of an epoch
as the sum of its actual cost and the recursive costs of the at most two induced epochs terminated by it as well as the at most two natural epochs
corresponding to them under the aforementioned mapping.

We henceforth change the constant 3 to 5 throughout the paper; in particular, instead of using $\log_3(n-1)$ levels, we will use $\log_5(n-1)$ levels.
Consequently, we allow the computation cost associated with any level-$\ell$ epoch to grow from $O(3^\ell)$ to  $O(5^\ell)$ (cf.\ Lemma \ref{basiccost}),
and obtain the recurrence $\hat C_{\ell} \le   4 \hat C_{\ell-1} + O(5^{\ell})$,
with the base condition
$\hat C_{0} =  O(1)$, which resolves to $\hat C_{\ell} = O(5^{\ell})$ (cf.\ Corollary \ref{important}).

\section{Improving the high probability bound} \label{app:improve}
If the length $t$ of the update sequence is $\Omega(n \log n)$, then our current high probability runtime bound  $O(t + n \log n)$ reduces   to $O(t)$.
It is natural to assume that $t \ge n/2$, otherwise some vertices are ``idle'', and we can simply ignore them.
(Such an assumption needs to be properly justified.
Nevertheless, dropping it usually triggers only minor adjustments.)
One may further assume that $t =  \Omega(n \log n)$, as this is indeed the case in many practical applications.
However, we believe that it is important to address the regime $n/2 \le t = o(n \log n)$, for both theoretical and practical reasons.

We do not make any assumption on $t$.
For $t = \Omega(n^\eps)$, we prove that the runtime exceeds $O(t)$ with probability (w.p.) polynomially small in $n$.
For smaller values of $t$, we  prove that the runtime exceeds $O(t)$ w.p.\ polynomially small in $t$.
Note that in the latter regime most of the $n$ vertices are idle,  and it does not make much sense that this probability would depend on the number of idle vertices;
nevertheless, for completeness, we show that the runtime exceeds $O(t + \log n \cdot \sqrt{t})$ w.p.\ polynomially small in $n$.

Let us revisit some of the details in the proof of Lemma \ref{expect}(2).
Recall that $A_\ell$ is the event that both $T_\ell > 4\log n$ and $T'_\ell \ge T_\ell / 4$  hold.
Instead, let us re-define $A_\ell$ to be the event that both $T_\ell > 4\log t$ and $T'_\ell \ge T_\ell / 4$  hold.
This modification triggers several changes.
First, Claim \ref{first1} will change to $\Prob(A_\ell) \le 8/ (3t^4)$.
Second, Claim \ref{negate} will be changed, so that if $\neg A_\ell$, then $Y_\ell < c(t_\ell + 3^\ell \cdot \log t)$.

In exactly the same way as before, we will be able to argue that $Y$ exceeds $O(t + n \log t)$ with probability $O(\log n / t^4)$.
However, this high probability bound is not what we are looking for.

The key insight for improving the high probability bound is given by the following lemma.
\begin{lemma} \label{maining}
The level of all vertices can be bounded by $\log_3 (2\sqrt{t})$, while increasing the total runtime of our algorithm by at most a constant factor.
\end{lemma}
Before proving this lemma, we demonstrate its power.
That is, we assume that the level of all vertices is bounded by $\log_3 (2\sqrt{t})$ and that the runtime required for that is negligible,
and show that the high probability bound can be improved under this assumption.
(Notice that we do not attempt to bound the out-degree  of vertices by $2\sqrt{t}$.)
In this case Equation (\ref{sumi}) in the proof of Corollary \ref{second2} will be changed to
$$Y ~=~ \sum_{\ell = 0}^{\log_3 (2\sqrt{t})} Y_\ell ~\le~ \sum_{\ell = 0}^{\log_3 (2\sqrt{t})} c(t_\ell + 3^\ell \cdot \log t)  ~\le~ c(t + (3/2) 2\sqrt{t} \log t).$$
Consequently, we can re-define $A$ to be the event that $Y > c(t + (3/2) 2\sqrt{t}\log t)$, without affecting the validity of Corollary \ref{second2}.
Finally, Equation (\ref{finall}) will be changed to
$$\Prob(A) ~\le~ \Prob(A_0 \cup A_1 \cup \ldots A_{\log_3 (2\sqrt{t})}) ~\le~ \sum_{\ell = 0}^{\log_3 (2\sqrt{t})} \Prob(A_\ell) ~\le~ (\log_3(2\sqrt{t}) + 1) (8/ (3t^4)) ~=~ O(\log t / t^4).$$
Thus, assuming the level of vertices is bounded by $\log_3 (2\sqrt{t})$,
$Y$ exceeds $c(t + (3/2) 2\sqrt{t}\log t) = O(t)$ w.p.\ $O(\log t/t^4)$, which is polynomially small in $t$ for all  $n$ and $t$
and polynomially small in $n$ for all $t = \Omega(n^\eps)$.

For the somewhat degenerate regime $t = o(n^{\eps})$, we simply use the original events $A_\ell$ and the original claims
(Claim \ref{first1} and \ref{negate}). Since the number of levels is bounded by $\log_3 (2\sqrt{t})$,
Equation (\ref{sumi}) in the proof of Corollary \ref{second2} will be changed to
$$Y ~=~ \sum_{\ell = 0}^{\log_3 (2\sqrt{t})} Y_\ell ~\le~ \sum_{\ell = 0}^{\log_3 (2\sqrt{t})} c(t_\ell + 3^\ell \cdot \log n)  ~\le~ c(t + (3/2) 2\sqrt{t} \log n).$$
Consequently, we can re-define $A$ to be the event that $Y > c(t + (3/2) 2\sqrt{t}\log n)$, without affecting the validity of Corollary \ref{second2}.
Finally, Equation (\ref{finall}) will be changed to
$$\Prob(A) ~\le~ \Prob(A_0 \cup A_1 \cup \ldots A_{\log_3 (2\sqrt{t})}) ~\le~ \sum_{\ell = 0}^{\log_3 (2\sqrt{t})} \Prob(A_\ell) ~\le~ (\log_3(2\sqrt{t}) + 1) (8/ (3n^4)) ~=~ O(\log t / n^4).$$
Thus, assuming the level of vertices is bounded by $\log_3 (2\sqrt{t})$,
$Y$ exceeds $c(t + (3/2) 2\sqrt{t}\log n) = O(t +   \log n \cdot \sqrt{t})$ w.p.\ $O(\log t/n^4)$, which is polynomially small in $n$ for all $t = o(n^\eps)$.
\vspace{10pt}
\noindent
\\
{\bf Proof of Lemma \ref{maining}:}
\vspace{3pt}
\\
Upon re-evaluating the level of a vertex, our algorithm may increase its level beyond $\ell_{max} := \log_3 (2\sqrt{t})$, as part of the rising process that creates
a new epoch of sufficiently high level, within Procedure $\rand$.
We adjust Procedure $\rand(v)$   by preventing $\ell^*$ from growing beyond $\ell_{max}$.
Specifically, we execute the loop in line 2 of the procedure as long as $\ell^* < \ell_{max}$, or in other words,
we adapt the continuation condition $\phi_v(\ell^*+1) \ge 3^{\ell^*+1}$ of the while loop to be $\phi_v(\ell^*+1) \ge 3^{\ell^*+1}$ and $\ell^* < \ell_{max}$.
As a result, we can no longer argue that the resulting level $\ell^*$ satisfies $\phi_v(\ell^*+1) < 3^{\ell^*+1}$.
While this upper bound on $\phi_v(\ell^*+1)$ may no longer hold,
note that the lower bound on $\phi_v(\ell^*)$, namely $\phi_v(\ell^*) \ge 3^{\ell^*}$, remains valid (cf.\ Lemma \ref{rand_basic}(3)).
The next observation thus follows from Corollary \ref{Ob:basic}.
\begin{observation} \label{obs:init}
Any vertex initiating a level-$\ell_{max}$ epoch has out-degree at least $3^{\ell_{max}} = 2\sqrt{t}$ at that time.
In particular, at least this number of edges incident on such a  vertex were inserted to the graph.
\end{observation}

As before, the mate $w$ of $v$ is chosen uniformly at random among $v$'s outgoing neighbors, and $\ell_w < \ell^*$.
More accurately, as detailed below, in some cases we restrict our attention to a \emph{subset} of $v$'s outgoing neighbors,
and choose the mate $w$ of $v$ uniformly at random among the vertices of this subset.
However, in any case, it is guaranteed that the level of $v$'s random mate $w$ will be strictly lower than that of $v$.

Another adjustment is to skip line 9 of Procedure $\rand(v)$ if $\ell^* = \ell_{max}$; indeed, if the new epoch is of
  level  $\ell_{max}$, we cannot make a recursive call that creates an epoch at   a higher level.


Note that the calls to $\setlvl(v,\ell^*)$ and $\setlvl(w,\ell^*)$ do not effect the incoming neighbors of $v$ and $w$ with maximum level $\ell_{max}$,
which   remain incoming to them after the calls.
Consequently, for any vertex of level $\ell_{max}$, all its outgoing edges do not flip.
Suppose that the new matched edge $(v,w)$ is of maximum level $\ell_{max}$, and consider the newly created epoch corresponding to it.
Since we guarantee that a random mate has a strictly lower level than the vertex choosing it,
the endpoints $v$ and $w$ of this epoch cannot be chosen as random mates of any vertex during the epoch's lifespan,
implying that such an epoch can be terminated only by deleting its associated edge $(v,w)$ from the graph.  
Thus any epoch of maximum level is a natural epoch, and all edges that are outgoing of its endpoints do not flip.
We have shown that, once a vertex rises to level $\ell_{max}$, the epoch associated with it becomes somewhat ``stagnant''.

Lemma \ref{rand_basic} implies that the runtime of the call to $\setlvl(v,\ell^*)$ (respectively, $\setlvl(w,\ell^*))$ is
$O(\dout(v) + \ell^*) = O(\dout(v))$ (resp., $O(\dout(w) + \ell^*) = O(\dout(w))$),
where $\dout(v) = \phi_v(\ell^*)$ (resp., $\dout(w) = \phi_w(\ell^*)$) is the new out-degree of $v$ (resp., $w$).
However, if $\ell^* = \ell_{max}$, we may not be able to upper bound neither $\dout(v)$ nor $\dout(w)$ by $O(3^{\ell^*}) = O(\sqrt{t})$.
Suppose w.l.o.g.\ that $\dout(v)  \ge \dout(w)$.
If $\dout(v) = O(3^{\ell^*})$, then our original analysis carries through.
We henceforth assume that $\ell^* = \ell_{max}$ and $\dout(v) := D \gg \sqrt{t}$, and show how to charge this $O(D)$ cost without creating an epoch at a higher level.


Consider the next time  that vertex $v$ becomes temporarily free, and denote by $\dout^{new}(v)$ the out-degree of $v$ at that time.
As mentioned, $v$ may become temporarily free only as a result of its matched edge $(v,w)$ being deleted from the graph, which terminates the corresponding level-$\ell_{max}$ epoch.
Since $v$ has become temporarily free, our update algorithm handles it by invoking Procedure $\evalf(v)$.

If $\dout^{new}(v) < 3^{\ell_{max}+1} = 6\sqrt{t}$,     Procedure $\evalf(v)$ calls to $\detr(v)$.
Due to the stagnation properties discussed above, all outgoing edges of $v$ at the time the epoch was created do not flip until its termination.
Hence, $D - \dout^{new}(v) = \Omega(D)$ edges incident on $v$  must have been deleted from the graph during this time interval,
and we can charge the original $O(D)$ cost to these edge deletions.

In the complementary case $\dout^{new}(v) \ge 3^{\ell_{max}+1} = 6\sqrt{t}$, this procedure calls to $\rand(v)$.
We make another adjustment to Procedure $\rand(v)$ for this particular case.
Specifically, in this case the procedure scans $3\sqrt{t}$ arbitrary outgoing neighbors of $v$,
and picks a mate $\tilde w$ for $v$ uniformly at random among $v$'s scanned neighbors that are of level strictly lower than $\ell_{max}$.
The level of $v$ remains $\ell_{max}$, and the level of $\tilde w$ is set to $\ell_{max}$ by calling   $\setlvl(\tilde w, \ell_{max})$, thus creating a level-$\ell_{max}$ epoch.
\begin{claim} \label{Obs:key}
For any vertex $v$,
at most $2\sqrt{t}$ of its neighbors may have level $\ell_{max}$ at any   point in time.
\end{claim}
\begin{proof}
Suppose for contradiction that more than $2\sqrt{t}$ neighbors of some vertex $v$ have level $\ell_{max}$ at some point in time.
Each of these neighbors is part of a single level-$\ell_{max}$ epoch at that time,
and the edges corresponding to these epochs are vertex-disjoint.
In particular, at least half of these neighbors of $v$ must have initiated a level-$\ell_{max}$ epoch,
or in other words, more than $\sqrt{t}$ vertices must have initiated a level-$\ell_{max}$-epoch.
Observation \ref{obs:init} implies that the total number of edges ever incident on those vertices exceeds $(\sqrt{t} \cdot 2\sqrt{t})/2 = t$,
contradicting the fact that the total number of edge updates is $t$.
\QED
\end{proof}
Claim \ref{Obs:key} implies
that $v$'s mate $\tilde w$ is chosen with probability at most $1/\sqrt{t}$. A key property is that the runtime $O(\sqrt{t})$ of Procedure $\rand(v)$ in this case does not depend on
$v$'s out-degree.

To summarize, the original $O(D)$ cost needed for rising a vertex $v$ to the maximum level $\ell_{max}$ is linear in the out-degree of $v$ at that time,
which may be prohibitively large.
However, the runtime of subsequent calls to Procedure $\rand(v)$ is $O(\sqrt{t})$, which is inverse-linear in the probability with which the matched edge is chosen,
and then our original analysis carries through. Since the level of $v$ remains $\ell_{max}$ in all such calls to Procedure $\rand(v)$,
the original $O(D)$ cost can be charged to the insertions of the $D$ outgoing edges of $v$ at that time.
On the other hand, if a subsequent call to Procedure $\detr(v)$ is made, then $v$'s level will decrease to $-1$ or 0,
which implies that we may have to spend an additional prohibitively large cost to rise $v$ to the maximum level $\ell_{max}$ in the future.
However, at least $\Omega(D)$ edge deletions incident on $v$ must have occurred until that time, to which we can charge the original $O(D)$ cost,
and then this charging argument can be reapplied from scratch.

This completes the proof of Lemma \ref{maining}. \QED

\section{Applications} \label{app:app}
{\bf Dynamic approximate MWMs.~}
As mentioned, Anand et al.\ \cite{ABGS12} gave a randomized algorithm for maintaining an 8-MWM in general $n$-vertex weighted graphs with expected update time $O(\log n \log \Delta)$.
Their algorithm maintains a partition of the edges in the graph into $O(\log \Delta)$ buckets according to their weight, with each bucket containing
edges of the same weight up to a constant factor. The maximal matching algorithm of \cite{BGS11} is employed (as a black-box) for each bucket separately.
By  carefully maintaining a matching in the graph obtained from the union of these $O(\log \Delta)$ maximal matchings, an 8-MWM is maintained in \cite{ABGS12}.
By plugging our improved algorithm, we   shave a factor of $\log n$ from the update time.
\begin{theorem}
Starting from an empty graph on $n$ fixed vertices, an 8-MWM can be maintained over any sequence of edge insertions and deletions
in expected amortized update time $O(\log \Delta)$, where $\Delta$ is the ratio between the maximum and minimum edge weights in the graph.
\end{theorem}
\vspace{4pt}
{\bf Distributed networks.~} Consider an arbitrary sequence of edge insertions and deletions in a distributed network.
Note that each vertex $v$ can gather complete information about its neighbors in two communication rounds.
Consequently, the na\"{\i}ve  (centralized) maximal matching algorithm discussed in Section 1.4 can be distributed in the obvious way,
requiring $O(1)$ communication rounds following a single edge update.
Moreover, messages of size $O(\log n)$ suffice for communicating the relevant information.
On the negative side, the number of messages sent per update may be as high as $O(n)$.
Note, however, that the total number of messages sent is upper bounded (up to a constant) by the total number of neighbor scans performed by the centralized algorithm.
This phenomenon extends far beyond the na\"{\i}ve maximal matching algorithm. In particular, it holds also w.r.t.\ Baswana et al.'s algorithm \cite{BGS11} and our algorithm.
Our analysis of Section \ref{sec:mainanal} shows that the amortized update time is constant,  
thus the average number of neighbor scans is also a  constant. (We did not try to optimize the latter constant, but it is rather small.)
\begin{theorem}
Starting from an empty distributed network on $n$ fixed vertices, a maximal matching (and thus 2-MCM and also 2-MCVC)
can be maintained distributively (under the $\mathcal{CONGEST}$ communication model) over any sequence of edge insertions and deletions with a constant amortized \emph{message complexity}.  
\end{theorem}
{\bf Remark.} Optimizing the constant behind the amortized message complexity is left as an open question.

\ignore{
\clearpage

\section{Junk}
[[S: Make sure to address the issue of level-1 matched edges]]
[[S: saving space: first paragraph of Sect. 2, the data structure section 3.1 (boring), Sect. 2.3.
Recall: I left the first paragraph in ``Baswana et al.'s scheme'' (in the technical overview section), can remove it.
Other places to save space: The footnote description on [AV-W14], the footnote on adversarial model.
Also: first sentence of ``wrapping up''. Maybe squeeze ack to one line.
Maybe can remove Lemma 3.4? don't think so... (but need to move backwards, to after Claim 3.2.
Should save space in applications. Remove some stuff from 1.3 (as it appears already in 1.4, and if not, then move there),
and then remove Organization part. I would also remove a line in Section 3.1]]







[[S: DEL: Also need to say that there are no edges between level-0 vertices, meaning that whenever the level of a vertex increases from -1 to 0 (turns from free to matched),
there is no violation, because all its incoming neighbors must be of level at least 0.]]

[[S: I don't mind having a too-large estimation on the level,
but I don't want to have a too-small estimation, and then to pick it randomly as a low-level vertex
only to find out (after evaluating its level) that its true level is larger than that.
(NOTE HOWEVER THAT THIS IS THE SITUATION ANYWAY... WE CAN'T CHANGE THIS)
If the estimation is too large, the only problem is to miss in this way a free vertex (whose level should be 0), but if we make sure to handle
those free vertices later (after evaluating their level as 0), then there is no problem.]]

[[S: choose one among $i$ and $\ell$ and be consistent, maybe add $j$ or other parameters]]

[[S: after Lemma 3.4]]
The key idea behind our charging scheme 
can be summarized as follows.
Consider a vertex $v$ that becomes temporarily free. If its out-degree is ``small'' (w.r.t.\ $3^{\ell_v}$), then we handle $v$ deterministically
and charge this cost to the past (i.e., to the termination cost of the corresponding level-$\ell_v$ epoch).
The interesting case is when the out-degree of $v$ is ``large'' (w.r.t.\ $3^{\ell_v}$), and then $v$ should rise to a higher level.
Since the outgoing neighbors of $v$ are of level at most $\ell_v$ and since we have complete information of $v$'s incoming neighbors,
we can compute the appropriate level $\ell^*$ to which $v$ should rise, and also flip the edges incident on $v$ accordingly, in time $O(3^{\ell^*})$; see Lemma \ref{rand_basic}(4).
Moreover, by Lemmas \ref{rand_basic}(2) and \ref{rand_basic}(3), after $v$ rises to level $\ell^*$, it has at least $3^{\ell^*}$ outgoing neighbors of level strictly lower than $\ell^*$.
Thus, by choosing one of them uniformly at random as a mate for $v$, the matched edge is chosen with probability at most $1/3^{\ell^*}$.
Consequently, we can charge this cost of $O(3^{\ell^*})$ to the future (i.e., to the creation cost of the corresponding level-$\ell^*$ epoch),
as the new epoch can cover this cost in the amortized sense.
}

\ignore{
\paragraph{important for talk}
It is difficult to orient edges according to out-degree. Because following a single edge flip, there may be a long and chaotic ``wave''.
What they suggested is to preserve the out-degree up to some small factor. This is still quite difficult.
What i suggest is different. I orient according to levels, but levels don't represent the out-degree. In this way I can be lazy.
Then when I need to handle a vertex of large out-degree, I simply raise its level, and pick at random a neighbor to pay for it.

The key idea is this -- if I scan a certain neighbor, then I will make sure, when choosing a random neighbor, to include it as one of the neighbors.

For the talk, think of a wave. It takes a logarithmic time to raise it up to a certain level, and then it fades away for free.
Instead, we don't raise the wave to any level, waves can grow without any control.

Oh, I disregarded all the issue of induced deletions. This is a pity. When choosing a random mate that is matched,
I need it to have a much lower out-degree

The algorithm of BGS11 partitions the vertices into $\log n$ levels, and the time needed to handle each level is constant,
which gives the $O(\log n)$ time. Maybe I can mention that it's not that just the average time is logarithmic, it is usually at least so due to
updating a logarithmic number of entries in the underlying array.
It is natural to work with $\log_k n$, for a larger $k$. Then the number of levels can be significantly reduced.
However, the time needed to handle each level will be $O(k)$, which gives an update time of $O(\log_k n \cdot k)$, optimized when $k = O(1)$.

I would say that the main difference is that BGS11 aims to rise to the highest possible level, constantly and persistently,
whereas we try to rise to the lowest possible level, and just at the time we handle the vertex. The fact that a vertex cannot rise to a higher level
(by their approach) immediately implies that the out-degree is small. But much more than that, it implies that whenever the level of a vertex falls
by one unit, the levels of its neighbors will not rise to a higher level. (When the level of a vertex rises, no other vertex is affected.)
So they get a fading wave phenomenon. But in order to do that, they have to spend a lot of time (and space) for determining the highest possible level.
Moreover, the fading wave phenomenon is somewhat complicated, in terms of implementation, but also conceptually.
As a result of our lazy approach, we can't say anything on the out-degree, except for immediately after evaluation. In fact, even immediately after evaluation,
we don't say anything (due to deterministic settle, which sets the level to 0 but don't flip the edges towards level-0 vertices). While we can overcome this issue,
there is absolutely no need. Whenever we handle a vertex, the time we spent can be covered either by the previous matched edge incident on it or the new one that we will next create. I think the key insight is that if a guy has large degree, then its out-degree may get bigger and bigger, but the bigger it gets the more tokens we should acquire,
and eventually we are prone to stop because the degree is at most $n-1$, hence the level cannot grow beyond $\log_3 (n-1)$.


Maybe there's a chance to get worst-case bounds:  Worst-case update time of $f(eps)$ (say $O(1/\eps)$), with an approximation factor w.h.p.\ $2+\eps$.
I don't know if this is worth the effort.
}
\end {document}